\documentclass[aps,prd,twocolumn,superscriptaddress,showpacs,preprintnumbers,amsmath,amssymb]{revtex4-1}

\usepackage[dvipdfmx]{graphicx}
\usepackage{dcolumn}
\usepackage{color}
\usepackage[mathlines]{lineno}
\usepackage{subfigure}

\graphicspath{{ps}}

\begin{document}

\preprint{\vbox{ \hbox{   }
    \hbox{Belle Preprint 2017-18}
    \hbox{KEK Preprint 2017-26}
}}

\title{Measurement of the $\tau$ lepton polarization and $R(D^*)$ in the decay $\bar{B} \rightarrow D^* \tau^- \bar{\nu}_\tau$\\
  with one-prong hadronic $\tau$ decays at Belle}

\noaffiliation
\affiliation{University of the Basque Country UPV/EHU, 48080 Bilbao}
\affiliation{Budker Institute of Nuclear Physics SB RAS, Novosibirsk 630090}
\affiliation{Faculty of Mathematics and Physics, Charles University, 121 16 Prague}
\affiliation{University of Cincinnati, Cincinnati, Ohio 45221}
\affiliation{Deutsches Elektronen--Synchrotron, 22607 Hamburg}
\affiliation{University of Florida, Gainesville, Florida 32611}
\affiliation{Department of Physics, Fu Jen Catholic University, Taipei 24205}
\affiliation{Justus-Liebig-Universit\"at Gie\ss{}en, 35392 Gie\ss{}en}
\affiliation{Gifu University, Gifu 501-1193}
\affiliation{SOKENDAI (The Graduate University for Advanced Studies), Hayama 240-0193}
\affiliation{Gyeongsang National University, Chinju 660-701}
\affiliation{Hanyang University, Seoul 133-791}
\affiliation{University of Hawaii, Honolulu, Hawaii 96822}
\affiliation{High Energy Accelerator Research Organization (KEK), Tsukuba 305-0801}
\affiliation{J-PARC Branch, KEK Theory Center, High Energy Accelerator Research Organization (KEK), Tsukuba 305-0801}
\affiliation{IKERBASQUE, Basque Foundation for Science, 48013 Bilbao}
\affiliation{Indian Institute of Science Education and Research Mohali, SAS Nagar, 140306}
\affiliation{Indian Institute of Technology Bhubaneswar, Satya Nagar 751007}
\affiliation{Indian Institute of Technology Guwahati, Assam 781039}
\affiliation{Indian Institute of Technology Hyderabad, Telangana 502285}
\affiliation{Indian Institute of Technology Madras, Chennai 600036}
\affiliation{Indiana University, Bloomington, Indiana 47408}
\affiliation{Institute of High Energy Physics, Chinese Academy of Sciences, Beijing 100049}
\affiliation{Institute of High Energy Physics, Vienna 1050}
\affiliation{Institute for High Energy Physics, Protvino 142281}
\affiliation{University of Mississippi, University, Mississippi 38677}
\affiliation{INFN - Sezione di Napoli, 80126 Napoli}
\affiliation{INFN - Sezione di Torino, 10125 Torino}
\affiliation{Advanced Science Research Center, Japan Atomic Energy Agency, Naka 319-1195}
\affiliation{J. Stefan Institute, 1000 Ljubljana}
\affiliation{Kanagawa University, Yokohama 221-8686}
\affiliation{Institut f\"ur Experimentelle Kernphysik, Karlsruher Institut f\"ur Technologie, 76131 Karlsruhe}
\affiliation{Kavli Institute for the Physics and Mathematics of the Universe (WPI), University of Tokyo, Kashiwa 277-8583}
\affiliation{Kennesaw State University, Kennesaw, Georgia 30144}
\affiliation{King Abdulaziz City for Science and Technology, Riyadh 11442}
\affiliation{Department of Physics, Faculty of Science, King Abdulaziz University, Jeddah 21589}
\affiliation{Korea Institute of Science and Technology Information, Daejeon 305-806}
\affiliation{Korea University, Seoul 136-713}
\affiliation{Kyoto University, Kyoto 606-8502}
\affiliation{Kyungpook National University, Daegu 702-701}
\affiliation{\'Ecole Polytechnique F\'ed\'erale de Lausanne (EPFL), Lausanne 1015}
\affiliation{P.N. Lebedev Physical Institute of the Russian Academy of Sciences, Moscow 119991}
\affiliation{Faculty of Mathematics and Physics, University of Ljubljana, 1000 Ljubljana}
\affiliation{Ludwig Maximilians University, 80539 Munich}
\affiliation{University of Malaya, 50603 Kuala Lumpur}
\affiliation{University of Maribor, 2000 Maribor}
\affiliation{Max-Planck-Institut f\"ur Physik, 80805 M\"unchen}
\affiliation{School of Physics, University of Melbourne, Victoria 3010}
\affiliation{University of Miyazaki, Miyazaki 889-2192}
\affiliation{Moscow Physical Engineering Institute, Moscow 115409}
\affiliation{Moscow Institute of Physics and Technology, Moscow Region 141700}
\affiliation{Graduate School of Science, Nagoya University, Nagoya 464-8602}
\affiliation{Kobayashi-Maskawa Institute, Nagoya University, Nagoya 464-8602}
\affiliation{Nara Women's University, Nara 630-8506}
\affiliation{National Central University, Chung-li 32054}
\affiliation{National United University, Miao Li 36003}
\affiliation{Department of Physics, National Taiwan University, Taipei 10617}
\affiliation{H. Niewodniczanski Institute of Nuclear Physics, Krakow 31-342}
\affiliation{Nippon Dental University, Niigata 951-8580}
\affiliation{Niigata University, Niigata 950-2181}
\affiliation{Novosibirsk State University, Novosibirsk 630090}
\affiliation{Osaka City University, Osaka 558-8585}
\affiliation{Pacific Northwest National Laboratory, Richland, Washington 99352}
\affiliation{Panjab University, Chandigarh 160014}
\affiliation{University of Pittsburgh, Pittsburgh, Pennsylvania 15260}
\affiliation{Punjab Agricultural University, Ludhiana 141004}
\affiliation{Theoretical Research Division, Nishina Center, RIKEN, Saitama 351-0198}
\affiliation{University of Science and Technology of China, Hefei 230026}
\affiliation{Showa Pharmaceutical University, Tokyo 194-8543}
\affiliation{Soongsil University, Seoul 156-743}
\affiliation{University of South Carolina, Columbia, South Carolina 29208}
\affiliation{Stefan Meyer Institute for Subatomic Physics, Vienna 1090}
\affiliation{Sungkyunkwan University, Suwon 440-746}
\affiliation{School of Physics, University of Sydney, New South Wales 2006}
\affiliation{Department of Physics, Faculty of Science, University of Tabuk, Tabuk 71451}
\affiliation{Tata Institute of Fundamental Research, Mumbai 400005}
\affiliation{Excellence Cluster Universe, Technische Universit\"at M\"unchen, 85748 Garching}
\affiliation{Department of Physics, Technische Universit\"at M\"unchen, 85748 Garching}
\affiliation{Toho University, Funabashi 274-8510}
\affiliation{Department of Physics, Tohoku University, Sendai 980-8578}
\affiliation{Department of Physics, University of Tokyo, Tokyo 113-0033}
\affiliation{Tokyo Institute of Technology, Tokyo 152-8550}
\affiliation{Tokyo Metropolitan University, Tokyo 192-0397}
\affiliation{University of Torino, 10124 Torino}
\affiliation{Utkal University, Bhubaneswar 751004}
\affiliation{Virginia Polytechnic Institute and State University, Blacksburg, Virginia 24061}
\affiliation{Wayne State University, Detroit, Michigan 48202}
\affiliation{Yamagata University, Yamagata 990-8560}
\affiliation{Yonsei University, Seoul 120-749}
  \author{S.~Hirose}\affiliation{Graduate School of Science, Nagoya University, Nagoya 464-8602} 
  \author{T.~Iijima}\affiliation{Kobayashi-Maskawa Institute, Nagoya University, Nagoya 464-8602}\affiliation{Graduate School of Science, Nagoya University, Nagoya 464-8602} 
  \author{I.~Adachi}\affiliation{High Energy Accelerator Research Organization (KEK), Tsukuba 305-0801}\affiliation{SOKENDAI (The Graduate University for Advanced Studies), Hayama 240-0193} 
  \author{K.~Adamczyk}\affiliation{H. Niewodniczanski Institute of Nuclear Physics, Krakow 31-342} 
  \author{H.~Aihara}\affiliation{Department of Physics, University of Tokyo, Tokyo 113-0033} 
  \author{S.~Al~Said}\affiliation{Department of Physics, Faculty of Science, University of Tabuk, Tabuk 71451}\affiliation{Department of Physics, Faculty of Science, King Abdulaziz University, Jeddah 21589} 
  \author{D.~M.~Asner}\affiliation{Pacific Northwest National Laboratory, Richland, Washington 99352} 
  \author{H.~Atmacan}\affiliation{University of South Carolina, Columbia, South Carolina 29208} 
  \author{T.~Aushev}\affiliation{Moscow Institute of Physics and Technology, Moscow Region 141700} 
  \author{R.~Ayad}\affiliation{Department of Physics, Faculty of Science, University of Tabuk, Tabuk 71451} 
% ...
\collaboration{The Belle Collaboration}

\begin{abstract}
  With the full data sample of $772 \times 10^6$ $B{\bar B}$ pairs recorded by the Belle detector at the KEKB electron-positron collider, the decay $\bar{B} \rightarrow D^* \tau^- \bar{\nu}_\tau$ is studied with the hadronic $\tau$ decays $\tau^- \rightarrow \pi^- \nu_\tau$ and $\tau^- \rightarrow \rho^- \nu_\tau$. The $\tau$ polarization $P_\tau(D^*)$ in two-body hadronic $\tau$ decays is measured, as well as the ratio of the branching fractions $R(D^{*}) = \mathcal{B}(\bar {B} \rightarrow D^* \tau^- \bar{\nu}_\tau) / \mathcal{B}(\bar{B} \rightarrow D^* \ell^- \bar{\nu}_\ell)$, where $\ell^-$ denotes an electron or a muon. Our results, $P_\tau(D^*) = -0.38 \pm 0.51 {\rm (stat)} ^{+0.21} _{-0.16} {\rm (syst)}$ and $R(D^*) = 0.270 \pm 0.035{\rm (stat)} ^{+0.028}_{-0.025}{\rm (syst)}$, are consistent with the theoretical predictions of the standard model. The polarization values of $P_\tau(D^*) > +0.5$ are excluded at the 90\% confidence level.
\end{abstract}

\pacs{13.20.He, 14.40.Nd}

\maketitle

\tighten

{\renewcommand{\thefootnote}{\fnsymbol{footnote}}}
\setcounter{footnote}{0}

\section{Introduction}

\begin{figure}[b!]
  \centering
  \includegraphics[width=7cm]{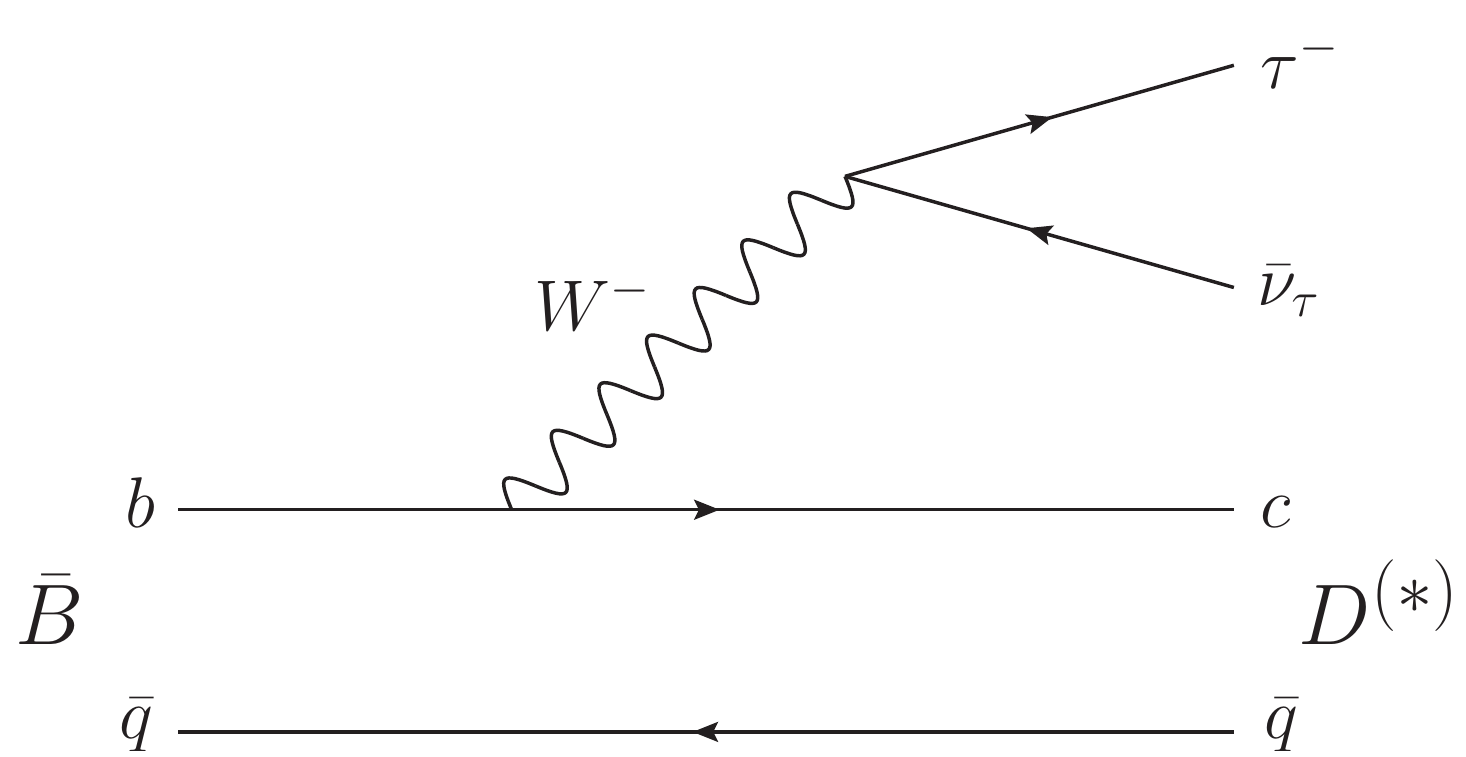}
  \caption{Feynman diagram of $\bar{B} \rightarrow D^{(*)} \tau^- \bar{\nu}_\tau$ for the SM amplitude, where $\bar{q}$ denotes $\bar{u}$ or $\bar{d}$.}
  \label{fig:Feynman}
\end{figure}

Semileptonic $B$ decays to $\tau$ leptons (semitauonic decays) are theoretically well-studied processes within the standard model (SM)~\cite{cite:Heiliger:1989, cite:Korner:1990, cite:Hwang:2000}, where the decay process is represented by the tree-level diagram shown in Fig.~\ref{fig:Feynman}. The $\tau$ lepton is more sensitive to new physics (NP) beyond the SM that couples strongly with mass. A prominent candidate is the two-Higgs-doublet model (2HDM)~\cite{cite:2HDM:1989}, where charged Higgs bosons appear. The contribution of the charged Higgs to the decay process $\bar{B} \to D^{(*)} \tau^- \bar{\nu}_\tau$~\cite{cite:CC} is suggested by many theoretical works (for example, Refs.~\cite{cite:Grzadkowski:1992,cite:Tanaka:1995,cite:Soni:1997,cite:Itoh:2005,cite:Crivellin:2012}).

Experimentally, the decays $\bar{B} \to D^{(*)} \tau^- \bar{\nu}_\tau$ have been studied by Belle~\cite{cite:Belle:2007, cite:Belle:2010, cite:Belle:2015, cite:Belle:2016}, \textit{BABAR}~\cite{cite:BaBar:2008, cite:BaBar:2012:letter, cite:BaBar:2013:fullpaper} and LHCb~\cite{cite:LHCb:2015}. Most of these studies have measured ratios of branching fractions, defined as
\begin{eqnarray}
  R(D^{(*)}) &=& \frac{\mathcal{B}(\bar{B} \to D^{(*)} \tau^- \bar{\nu}_\tau)}{\mathcal{B}(\bar{B} \to D^{(*)} \ell^- \bar{\nu}_\ell)}.
\end{eqnarray}
The denominator is the average of $\ell^- = e^-,~\mu^-$ for Belle and \textit{BABAR}, and $\ell^- = \mu^-$ for LHCb. The ratio cancels numerous uncertainties common to the numerator and the denominator; these include the uncertainty in the Cabibbo-Kobayashi-Maskawa matrix element $|V_{cb}|$, many of the theoretical uncertainties on hadronic form factors (FFs), and experimental reconstruction effects. Recently, LHCb measured the $\bar{B}^0 \rightarrow D^{*+} \tau^- \bar{\nu}_\tau$ mode using the three-prong decay $\tau^- \rightarrow \pi^- \pi^+ \pi^- (\pi^0) \nu_\tau$~\cite{cite:LHCb:2017}. To reduce the systematic uncertainty, $R_{\rm had}(D^*) = \mathcal{B}(\bar{B}^0 \to D^{*+} \tau^- \bar{\nu}_\tau) / \mathcal{B}(\bar{B}^0 \to D^{*+} \pi^- \pi^+ \pi^-)$ is measured with the common final states between the numerator and the denominator, and $R_{\rm had}(D^*)$ is converted to $R(D^*)$ by using the world-average values for $\mathcal{B}(\bar{B}^0 \to D^{*+} \pi^- \pi^+ \pi^-)$ and $\mathcal{B}(\bar{B}^0 \to D^{*+} \mu^- \bar{\nu}_\mu)$.

As of early 2016, the results from the three experiments~\cite{cite:Belle:2015,cite:Belle:2016,cite:BaBar:2012:letter,cite:BaBar:2013:fullpaper,cite:LHCb:2015} were $1.9$ and $3.3$ standard deviations ($\sigma$)~\cite{cite:HFLAV:2014} away from the SM predictions of $R(D) = 0.299 \pm 0.011$~\cite{cite:RD_FermiandMILC:2015} or $0.300 \pm 0.008$~\cite{cite:RD_HPQCD:2015} and $R(D^*) = 0.252 \pm 0.003$~\cite{cite:RDst:2012}, respectively. The overall discrepancy with the SM was about 4$\sigma$. These deviations have been theoretically studied in the context of various NP models~\cite{cite:RDst:2012, cite:Datta:2012, cite:Celis:2013, cite:Tanaka:2013, cite:Biancofiore:2013, cite:Dorsner:2013, cite:Sakaki:2013, cite:Hagiwara:2014, cite:Duraisamy:2014, cite:Sakaki:2015, cite:Freytsis:2015, cite:Li:2016, cite:Bhattacharya:2016, cite:Bardhan:2017, cite:Celis:2017}.

In addition to $R(D^{(*)})$, the polarizations of the $\tau$ lepton and the $D^*$ meson are sensitive to NP~\cite{cite:Tanaka:1995, cite:Tanaka:2010, cite:RDst:2012, cite:Datta:2012, cite:Biancofiore:2013, cite:Tanaka:2013, cite:Sakaki:2013, cite:Duraisamy:2014, cite:Bhattacharya:2016, cite:Bardhan:2017}. The $\tau$ lepton polarization is defined as
\begin{eqnarray}
  P_\tau(D^{(*)}) &=& \frac{\Gamma^+(D^{(*)}) - \Gamma^-(D^{(*)})}{\Gamma^+(D^{(*)}) + \Gamma^-(D^{(*)})},
\end{eqnarray}
where $\Gamma^{\pm}(D^{(*)})$ denotes the decay rate of $\bar{B} \rightarrow D^{(*)} \tau^- \bar{\nu}_\tau$ with a $\tau$ helicity of $\pm 1/2$. The SM predicts $P_\tau(D) = 0.325 \pm 0.009$~\cite{cite:Tanaka:2010} and $P_\tau(D^*) = -0.497 \pm 0.013$~\cite{cite:Tanaka:2013}. For example, the type-II 2HDM allows $P_\tau(D^{(*)})$ to be between $-0.6$ and $+1.0$ for $\bar{B} \rightarrow D \tau^- \bar{\nu}_\tau$ and between $-0.7$ and $+1.0$ for $\bar{B} \rightarrow D^* \tau^- \bar{\nu}_\tau$~\cite{cite:Tanaka:2013,cite:comment:2HDMII}, whereas a leptoquark model suggested in Ref.~\cite{cite:Sakaki:2013} with a leptoquark mass of 1~TeV allows $P_\tau(D^*)$ to be between $-0.5$ and $0.0$. $P_\tau(D^{(*)})$ can be measured in two-body hadronic $\tau$ decays with the differential decay rate
\begin{eqnarray}
  \frac{1}{\Gamma(D^{(*)})} \frac{d \Gamma(D^{(*)})}{d \cos\theta_{\rm hel}} &=& \frac{1}{2}\left[1 + \alpha P_\tau(D^{(*)}) \cos \theta_{\rm hel} \right],\label{eq:coshel}
\end{eqnarray}
where $\theta_{\rm hel}$ is the angle of the $\tau$-daughter meson momentum with respect to the direction opposite the momentum of the $\tau^- \bar{\nu}_\tau$ system in the rest frame of $\tau$. The parameter $\alpha$ describes the sensitivity to $P_\tau(D^{(*)})$ for each $\tau$-decay mode; in particular, $\alpha = 1$ for $\tau^- \rightarrow \pi^- \nu_\tau$ and $\alpha = 0.45$ for $\tau^- \rightarrow \rho^- \nu_{\tau}$~\cite{cite:Hagiwara:1990}.

In this paper, we describe details of the first $P_\tau(D^*)$ measurement in the decay $\bar{B} \rightarrow D^* \tau^- \bar{\nu}_\tau$ with the $\tau$ decays $\tau^- \rightarrow \pi^- \nu_{\tau}$ and $\tau^- \rightarrow \rho^- \nu_{\tau}$ reported in Ref.~\cite{cite:Belle:2017}. Our study includes an $R(D^*)$ measurement independent of previous studies~\cite{cite:Belle:2015, cite:Belle:2016, cite:BaBar:2012:letter, cite:BaBar:2013:fullpaper, cite:LHCb:2015}, in which leptonic $\tau$ decays have been used.

\section{Experimental Apparatus}

We use the full $\Upsilon(4S)$ data sample containing $772 \times 10^6 B\bar{B}$ pairs recorded with the Belle detector~\cite{cite:Belle-detector:2002} at the asymmetric-beam-energy $e^+ e^-$ collider KEKB~\cite{cite:KEKB:2003}. The Belle detector is a large-solid-angle magnetic spectrometer that consists of a silicon vertex detector (SVD), a 50-layer central drift chamber (CDC), an array of aerogel threshold Cherenkov counters (ACC), a barrel-like arrangement of time-of-flight scintillation counters (TOF) and an electromagnetic calorimeter (ECL) comprised of CsI(Tl) crystals located inside a superconducting solenoid coil that provides a 1.5~T magnetic field. An iron flux-return located outside the coil is instrumented to detect $K_L^0$ mesons and to identify muons (KLM). The detector is described in detail elsewhere~\cite{cite:Belle-detector:2002}. Two inner detector configurations were used. A 2.0~cm radius beampipe and a 3-layer SVD were used for the first sample of $152 \times 10^6 B\bar{B}$ pairs, while a 1.5~cm radius beampipe, a 4-layer SVD and a small-cell inner drift chamber were used to record the remaining $620 \times 10^6 B\bar{B}$ pairs~\cite{cite:SVD2:2006}.

\section{Monte Carlo Simulation}

The Monte Carlo (MC) simulated events are used to establish the analysis criteria, study the background and estimate the signal reconstruction efficiency. Events with a $B \bar{B}$ pair are generated using \textsc{EvtGen}~\cite{cite:EvtGen:2001}, and the $B$ meson decays are reproduced based on branching fractions reported in Ref.~\cite{cite:PDG:2016}. The hadronization process of the $B$ meson decay with no experimentally-measured branching fraction is inclusively reproduced by \textsc{Pythia}~\cite{cite:PYTHIA:2006}. For continuum $e^+ e^- \rightarrow q{\bar q}$ ($q = u, d, s, c$) events, hadronization of the initial quark pair is described by \textsc{Pythia}, and hadron decays are modeled by \textsc{EvtGen}. Final-state radiation from charged particles is added using \textsc{Photos}~\cite{cite:PHOTOS:2016}. Detector responses are reproduced by the Belle detector simulator based on \textsc{Geant}3~\cite{cite:GEANT:1984}. The MC samples used in this analysis are described below.

\begin{description}

\item[\textbf{\boldmath${\bar B} \rightarrow D^* \tau^- \bar{\nu}_\tau$}]\mbox{}\\
  The MC sample for the signal mode (${\bar B} \rightarrow D^* \tau^- \bar{\nu}_\tau$) is generated with hadronic FFs based on heavy quark effective theory (HQET). The following values of the hadronic FF parameters in the Caprini-Lellouch-Neubert scheme~\cite{cite:Caprini:1998, cite:CLN:note} are used: $\rho^2 = 1.207 \pm 0.015 \pm 0.021$, $R_1 = 1.403 \pm 0.033$, and $R_2 = 0.854 \pm 0.020$ from the experimental world averages~\cite{cite:HFLAV:2014}, and $R_0 = 1.22$ with 10\% uncertainty from the HQET estimation~\cite{cite:RDst:2012}.

\item[\textbf{\boldmath${\bar B} \rightarrow D^* \ell^- \bar{\nu}_\ell$}]\mbox{}\\
  The MC sample for the normalization mode ($\bar{B} \rightarrow D^* \ell^- \bar{\nu}_\ell$) is generated based on HQET. Since the FF parameters used for the production of the normalization MC sample have been updated as described above, final-state kinematics are corrected to match the latest parameter values.
  
\item[\textbf{\boldmath${\bar B} \rightarrow D^{**} \ell^- \bar{\nu}_\ell$} and \textbf{\boldmath${\bar B} \rightarrow D^{**} \tau^- \bar{\nu}_\tau$}]\mbox{}\\
  Semileptonic decays ${\bar B} \rightarrow D^{**} \ell^- {\bar \nu_{\ell}}$ and ${\bar B} \rightarrow D^{**} \tau^- {\bar \nu_{\tau}}$, where $D^{**}$ denotes the excited charm meson states heavier than $D^*$, comprise an important background category as they have a similar decay topology to the signal events. The MC sample for ${\bar B} \rightarrow D^{**} \ell^- \bar{\nu}_\ell$ is generated based on the Isgur-Scora-Grinstein-Wise (ISGW) model~\cite{cite:ISGW2:1995}, and decay kinematics are corrected to match the Leibovich-Ligeti-Stewart-Wise (LLSW) model~\cite{cite:LLSW:1998}. The branching fractions for $\bar{B} \rightarrow D^{**} \ell^- \bar{\nu}_\ell$ with $D^{**} = D_0^*$, $D_1$, $D'_1$ and $D_2^*$ are taken from the world averages~\cite{cite:HFLAV:2014}. For the $D^{**}$ decays, in addition to experimentally-measured modes, we allow unmeasured final states consisting of a $D^{(*)}$ and one or two pions, a $\rho$ meson, or an $\eta$ meson based on quantum-number, phase-space and isospin considerations. The radially-excited $D^{(*)}(2S)$ modes are included so that the total branching fraction of $\bar{B} \rightarrow D^{**} \ell^- \bar{\nu}_\ell$ becomes about 3\%, which is expected from the difference between $\mathcal{B}(\bar{B} \rightarrow X_c \ell^- \bar{\nu}_\ell)$ (where $X_c$ denotes all the possible charmed-meson states) and the sum of the exclusive branching fractions of $\mathcal{B}(\bar{B} \rightarrow D^{(*)} \ell^- \bar{\nu}_\ell)$. The ${\bar B} \rightarrow D^{**} \tau^- \bar{\nu}_\tau$ MC sample is generated using the ISGW model. We take the branching fractions from the theoretical estimates of $R(D^{**}) \equiv \mathcal{B}(\bar{B} \rightarrow D^{**} \tau^- \bar{\nu}_\tau) / \mathcal{B}(\bar{B} \rightarrow D^{**} \ell^- \bar{\nu}_\ell)$ for each $D^{**}$ state~\cite{cite:RDstst:2017}. We use the average $R(D^{**})$ of the four approximations discussed in Ref~\cite{cite:RDstst:2017}. We do not consider $\bar{B} \rightarrow D^{(*)}(2S) \tau^- \bar{\nu}_\tau$ or other semitauonic modes containing a charmed state heavier than $D^{(*)}(2S)$ as their small phase space suppresses the branching fractions. 

\item[Other background]\mbox{}\\
  The MC samples for other background processes, both $B \bar{B}$ events and continuum $e^+ e^- \rightarrow q{\bar q}$ events, are generated based on the past experimental studies reported in Ref.~\cite{cite:PDG:2016}. Unmeasured decay channels are generated with \textsc{Pythia} through the inclusive hadronization process.
\end{description}

The MC sample sizes of the signal mode, the normalization mode, ${\bar B} \rightarrow D^{**} \ell^- \bar{\nu}_\ell$, ${\bar B} \rightarrow D^{**} \tau^- \bar{\nu}_\tau$, the $B \bar{B}$ background, and the $q \bar{q}$ process are 40, 10, 40, 400, 10, and 5 times larger, respectively, than the full Belle data sample.

\section{Event Reconstruction}

\subsection{Reconstruction of the tag side}\label{sec:recon-tagside}

We conduct the analysis by first identifying events where one of the two $B$ mesons ($B_{\rm tag}$) is reconstructed in one of 1104 exclusive hadronic $B$ decays~\cite{cite:Full-recon:2011}. A hierarchical multivariate algorithm based on the NeuroBayes neural-network package is employed. More than 100 input variables are used to determine well-reconstructed $B$ candidates, including the difference between the energy of the reconstructed $B_{\rm tag}$ candidate and the beam energy in the $e^+ e^-$ center-of-mass (CM) frame $\Delta E \equiv E_{\rm tag}^* - E_{\rm beam}^*$, as well as the event shape variables for suppression of $e^+ e^- \rightarrow q{\bar q}$ background. The quality of the $B_{\rm tag}$ candidate is synthesized in a single NeuroBayes output-variable classifier ($O_{\rm NB}$). We require the beam-energy-constrained mass of the $B_{\rm tag}$ candidate $M_{\rm bc} \equiv \sqrt{ E_{\rm beam}^{*2} - |\vec{p}_{\rm tag}^{\kern2pt *}|^2}$, where $\vec{p}_{\rm tag}^{\kern2pt *}$ is the reconstructed $B_{\rm tag}$ three-momentum in the CM frame, to be greater than 5.272~GeV and the value of $\Delta E$ to be between $-150$ and $100~{\rm MeV}$. Throughout the paper, natural units with $\hbar = c = 1$ are used. We place a requirement on $O_{\rm NB}$ such that about 90\% of true $B_{\rm tag}$ and about 30\% of fake $B_{\rm tag}$ candidates are retained. If two or more $B_{\rm tag}$ candidates are retained in one event, we select the one with the highest $O_{\rm NB}$.

Due to limited knowledge of hadronic $B$ decays, the branching fractions of the $B_{\rm tag}$ decay modes are not perfectly modeled in the MC simulation. It is therefore essential to calibrate the $B_{\rm tag}$ reconstruction efficiency (tagging efficiency) with control data samples. We determine a scale factor for each $B_{\rm tag}$ decay mode using events where the signal-side $B$ meson candidate ($B_{\rm sig}$) is reconstructed in $\bar{B} \rightarrow D^{(*)} \ell^- \bar{\nu}_\ell$ modes. Further details of the calibration method are described in Ref.~\cite{cite:Belle_Xulnu:2013}. The ratio of measured to expected rates in each decay mode ranges from 0.2 to 1.4, depending on the $B_{\rm tag}$ decay mode, and is 0.72 on average. After the efficiency calibration, the tagging efficiencies are estimated to be about 0.20\% for charged $B$ mesons and 0.15\% for neutral $B$ mesons.

\subsection{Reconstruction of the signal side}\label{sec:event-reconstruction}

We reconstruct the signal mode and the normalization mode using the particle candidates not used for $B_{\rm tag}$ reconstruction. The following decay modes are used for the $B_{\rm sig}$ daughter particles: $D^{*0} \rightarrow D^0 \gamma$, $D^0 \pi^0$, $D^{*+} \rightarrow D^+ \pi^0$, and $D^0 \pi^+$ for the $D^*$ candidate; $\tau^- \rightarrow \pi^- \nu_\tau$ and $\rho^- \nu_\tau$ for the $\tau$ candidate; $D^0 \rightarrow K_S^0 \pi^0$, $\pi^+ \pi^-$, $K^- \pi^+$, $K^+ K^-$, $K^- \pi^+ \pi^0$, $K_S^0 \pi^+ \pi^-$, $K_S^0 \pi^+ \pi^- \pi^0$, $K^- \pi^+ \pi^+ \pi^-$, $D^+ \rightarrow K_S^0 \pi^+$, $K_S^0 K^+$, $K_S^0 \pi^+ \pi^0$, $K^- \pi^+ \pi^+$, $K^+ K^- \pi^+$, $K^- \pi^+ \pi^+ \pi^0$, and $K_S^0 \pi^+ \pi^+ \pi^-$ for the $D$ candidate; and $K_S^0 \rightarrow \pi^+ \pi^-$, $\pi^0 \rightarrow \gamma \gamma$ and $\rho^- \rightarrow \pi^- \pi^0$, respectively, for the light-meson candidates. A $\tau$-daughter candidate $\pi^-$ or $\rho^-$ is combined with a $D^*$ candidate to form a $B_{\rm sig}$ candidate. For the normalization events, a charged lepton $e^-$ or $\mu^-$ is associated instead of $\pi^-$ or $\rho^-$.

\subsubsection{Particle selection}

First, daughter particles of $D^*$ and $\tau$ ($K^\pm$, $\pi^\pm$, $K_S^0$, $\gamma$, $\pi^0$ and $\rho^\pm$) and charged leptons ($e^\pm$ and $\mu^\pm$) are reconstructed. For $B_{\rm sig}$ reconstruction, we use different particle selections from those applied for the $B_{\rm tag}$ reconstruction described in Ref.~\cite{cite:Full-recon:2011}. 

Charged particles are reconstructed using the SVD and the CDC. All tracks, except for $K_S^0$-daughter candidates, are required to have $dr < 0.5$~cm and $|dz| < 2.0$~cm, where $dr$ and $|dz|$ are the impact parameters to the interaction point (IP) in the directions perpendicular and parallel, respectively, to the $e^+$ beam axis. Charged-particle types are identified by a likelihood ratio based on the responses of the sub-detector systems. Identification of $K^{\pm}$ and $\pi^{\pm}$ candidates is performed by combining measurements of specific ionization ($dE/dx$) in the CDC, the time of flight from the IP to the TOF counter and the photon yield in the ACC. For $\tau$-daughter $\pi^{\pm}$ candidates, an additional proton veto is required in order to reduce background from baryonic $B$ decays such as ${\bar B} \rightarrow D^* {\bar p} n$. The ECL electromagnetic shower shape, track-to-cluster matching at the inner surface of the ECL, $dE/dx$ in the CDC, the photon yield in the ACC and the ratio of the cluster energy in the ECL to the track momentum measured with the SVD and the CDC are used to identify $e^{\pm}$ candidates~\cite{cite:eID:2002}. Muon candidates are selected based on their penetration range and transverse scattering in the KLM~\cite{cite:muID:2002}. To form $K_S^0$ candidates, we combine pairs of oppositely-charged tracks, treated as pions. Standard Belle $K_S^0$ selection criteria are applied~\cite{cite:Belle_KsKsKs:2005}: the reconstructed vertex must be detached from the IP, the momentum vector must point back to the IP, and the invariant mass must be within $\pm$30~MeV of the nominal $K_S^0$ mass~\cite{cite:PDG:2016}, which corresponds to about 8$\sigma$. (In this section, $\sigma$ denotes the corresponding mass resolution.)

Photons are reconstructed using ECL clusters not matching to charged tracks. Photon energy thresholds of 50, 100 and 150~MeV are used in the barrel, forward-endcap and backward-endcap regions, respectively, of the ECL to reject low-energy background photons, such as those originating from the $e^+ e^-$ beams and hadronic interactions of particles with materials in the detector.

Neutral pions are reconstructed in the decay $\pi^0 \to \gamma\gamma$. For $\pi^0$ candidates from $D$ or $\rho$ decay, referred to as normal $\pi^0$s, we impose the same photon energy thresholds described above. The $\pi^0$ candidate's invariant mass must lie between 115 and 150 MeV, corresponding to about $\pm 3\sigma$ around the nominal $\pi^0$ mass~\cite{cite:PDG:2016}. In order to reduce the number of fake $\pi^0$ candidates, we apply the following $\pi^0$ candidate selection procedure. The $\pi^0$ candidates are sorted in descending order according to the energy of the most energetic daughter. If a given photon is the most energetic daughter of two or more candidates, they are sorted by the energy of the lower-energy daughter. We then retain the $\pi^0$ candidates whose daughter photons are not shared with a higher-ranked candidate. In this criterion, 76\% of the correctly reconstructed $\pi^0$ candidates are selected while 54\% of the fake $\pi^0$ candidates are removed. The retained $\pi^0$ candidates are used for $D$ and $\rho$ reconstruction described later.

For the soft $\pi^0$ from $D^*$ decay, we impose a relaxed photon energy threshold of 22 MeV in all ECL regions and the same requirement for the invariant mass of the two photons. Additionally, the energy asymmetry $A_{\pi^0} = (E_h - E_l)/(E_h + E_l)$ is required to be less than 0.6, where $E_h$ and $E_l$ are the energies of the high- and low-energy photon daughters in the laboratory frame. Here, we do not apply the normal-$\pi^0$ candidate selection procedure.

The $\rho$ candidate is formed from the combination of a $\pi^{\pm}$ and a $\pi^0$. The candidate invariant mass must lie between 0.66 and 0.96~GeV.

\subsubsection{$D^{(*)}$ reconstruction}\label{sec:recon-Dst-recon}

After reconstructing the light mesons, we reconstruct the $D$ candidates in 15 decay modes. The $D$ invariant mass requirements are optimized for each decay mode. For the $D^0$ modes used in forming $D^{*0}$ candidates, the reconstructed invariant masses ($M_D$) are required to be within $\pm 2.0\sigma$ ($\pm 1.5\sigma$) of the nominal $D^0$ meson mass~\cite{cite:PDG:2016} for the high (low) signal-to-noise ratio (SNR) modes. For $D^{*+} \rightarrow D^0 \pi^+$ candidates, the $M_D$ requirements are loosened to $\pm 4.0\sigma$ and $\pm 2.0\sigma$ for the high- and low-SNR modes, respectively. The requirements for the $D^+$ candidates are $\pm 2.5\sigma$ for the high-SNR modes and $\pm 1.5\sigma$ for the low-SNR modes around the nominal $D^+$ meson mass~\cite{cite:PDG:2016}. Here, the high-SNR modes are $D^0 \rightarrow K_S^0 \pi^0$, $K^- \pi^+$, $K^+ K^-$, $K_S^0 \pi^+ \pi^-$, $K^- \pi^+ \pi^+ \pi^-$, $D^+ \rightarrow K_S^0 \pi^+$, $K_S^0 K^+$, $K^- \pi^+ \pi^+$; the low-SNR modes are all remaining $D$ modes. We reconstruct $D^*$ candidates by combining a $D$ candidate with a $\pi^{\pm}$, $\gamma$, or soft $\pi^0$. The $D^*$ candidates are selected based on the mass difference $\Delta M \equiv M_{D^*} - M_D$, where $M_{D^*}$ denotes the reconstructed invariant mass of the $D^*$ candidate. The $D^{*0} \rightarrow D^0 \gamma$, $D^{*0} \rightarrow D^0 \pi^0$, $D^{*+} \rightarrow D^+ \pi^0$, and $D^{*+} \rightarrow D^0 \pi^+$ candidates are required to have $\Delta M$ within $\pm 1.5\sigma$, $\pm 2.0\sigma$, $\pm 2.0\sigma$ and $\pm 3.5\sigma$, respectively, of the nominal $\Delta M$.

\subsubsection{$B_{\rm sig}$ selection}

The $B_{\rm sig}$ candidates are formed by associating a $\tau$-daughter meson (signal events) or a $\ell^-$ (normalization events) with a $D^*$ candidate. Allowed combinations are $D^{*-} + d^+$ for $B_{\rm sig}^0$, $D^{*+} + d^-$ for $\bar{B}_{\rm sig}^0$, $\bar{D}^{*0} + d^+$ for $B_{\rm sig}^+$ and $D^{*0} + d^-$ for $B_{\rm sig}^-$, where $d^- = \pi^-, \rho^-$ or $\ell^-$.  We select one of the following $B$ meson combinations: $(B_{\rm sig}^0, \bar{B}_{\rm tag}^0)$, $(\bar{B}_{\rm sig}^0, B_{\rm tag}^0)$, $(B_{\rm sig}^+, B_{\rm tag}^-)$ and $(B_{\rm sig}^-, B_{\rm tag}^+)$.

\begin{figure}[t!]
  \centering
  \includegraphics[width=9cm]{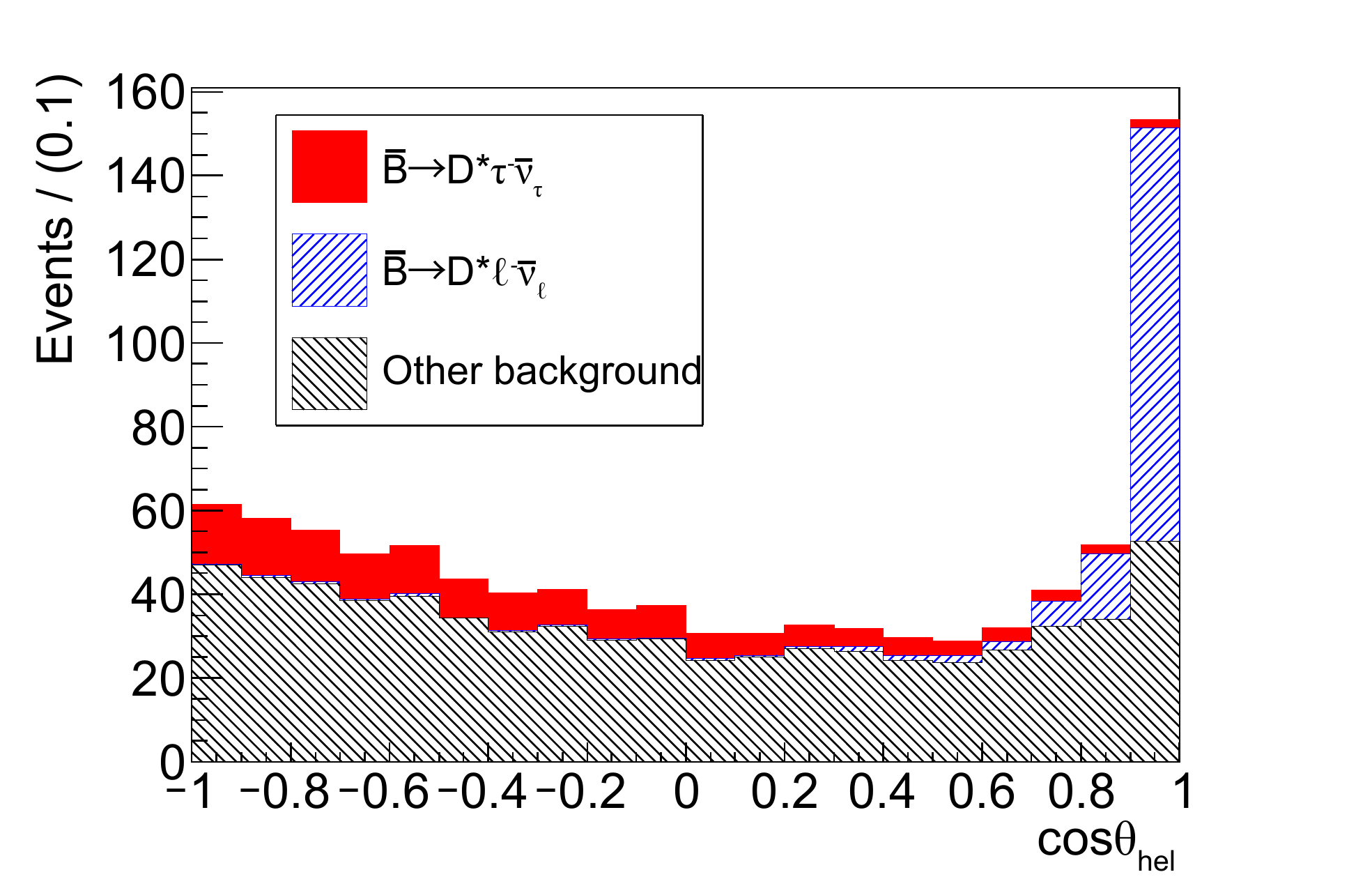}
  \caption{Distribution of $\cos\theta_{\rm hel}$ (MC) for the signal (red), $\bar{B} \rightarrow D^* \ell^- \bar{\nu}_\ell$ (blue-hatched) and the other background (black-hatched) in the $\tau^- \rightarrow \pi^- \nu_\tau$ sample. The SM prediction on $P_\tau(D^*)$ is assumed. All the signal selection requirements including $E_{\rm ECL}$ and $q^2$ are applied.}
  \label{fig:coshel_peak}
\end{figure}

For the signal mode, if at least one possible candidate for the signal mode is found in an event, we calculate $\cos \theta_{\rm hel}$ in the rest frame of the $\tau$. Although this frame cannot be determined completely, equivalent kinematic information is obtained using the rest frame of the $\tau^- \bar{\nu}_\tau$ system. This frame is obtained by boosting the laboratory frame along with the three-momentum vector component of the momentum transfer
\begin{eqnarray}
  q &=& p_{e^+ e^-} - p_{\rm tag} - p_{D^*},
\end{eqnarray}
where $p$ denotes the four-momentum of the $e^+ e^-$ beam, $B_{\rm tag}$, and $D^*$, respectively. In this frame, the energy and the magnitude of the momentum of the $\tau$ lepton are determined only by $q^2$ as
\begin{eqnarray}
  E_\tau           &=& \frac{q^2 + m_{\tau}^2}{2 \sqrt{q^2}},\\
  |\vec{p}_{\tau}| &=& \frac{q^2 - m_{\tau}^2}{2 \sqrt{q^2}},
\end{eqnarray}
where $m_\tau$ is the $\tau$ lepton mass. The cosine of the angle between the momenta of the $\tau$ lepton and its daughter meson is determined by
\begin{eqnarray}
  \cos \theta_{\tau d} &=& \frac{2 E_\tau E_d - m_{\tau}^2 - m_d^2} {2 |\vec{p}_{\tau}||\vec{p}_d|},
\end{eqnarray}
where $E_{\tau(d)}$ and $\vec{p}_{\tau(d)}$ denote the energy and the momentum of the $\tau$ lepton (the $\tau$ daughter $d$) respectively, and $m_d$ is the mass of the $\tau$ daughter. Through a Lorentz transformation from the rest frame of the $\tau^- \bar{\nu}_\tau$ system to the $\tau$ rest frame, the following relation is obtained:
\begin{eqnarray}
  |\vec{p}_d^{\kern2pt \tau}| \cos\theta_{\rm hel} &=& -\gamma |\vec{\beta}| E_d + \gamma |\vec{p}_d| \cos\theta_{\tau d},
\end{eqnarray}
where $|\vec{p}_d^{\kern2pt \tau}| = (m_\tau^2 - m_d^2) / 2 m_\tau$ is the $\tau$-daughter momentum in the rest frame of $\tau$, and $\gamma = E_\tau / m_\tau$ and $|\vec{\beta}| = |\vec{p}_\tau| / E_\tau$. Solving gives the value of $\cos \theta_{\rm hel}$. Events are required to lie in the physical region of $|\cos\theta_{\rm hel}| < 1$, where 97\% of the reconstructed signal events are retained. As shown in Fig.~\ref{fig:coshel_peak}, there is a significant background peak near 1 in the $\tau^- \rightarrow \pi^- \nu_\tau$ sample due to the $\bar{B} \rightarrow D^* \ell^- \bar{\nu}_\ell$ background. To reject this background, we only use the region $\cos\theta_{\rm hel} < 0.8$ in the fit to the $\tau^- \rightarrow \pi^- \nu_\tau$ sample.

Due to the kinematic constraint that $q^2$ must be greater than $m_\tau^2$, almost no signal events exist with $q^2$ below 4~GeV$^2$. Therefore $q^2 > 4~{\rm GeV}^2$ is required. The variable $E_{\rm ECL}$ is the linear sum of the energy of ECL clusters not used in the event reconstruction. The ECL clusters satisfying the photon-energy requirement defined in the previous section are added to $E_{\rm ECL}$. Signal events ideally have $E_{\rm ECL}$ equal to zero with a tail in the $E_{\rm ECL}$ distribution from the beam background and split-off showers, separated from the main ECL cluster and reconstructed as photon candidates. We require $E_{\rm ECL}$ to be less than 1.5~GeV. 

For the normalization mode, we calculate the squared missing mass,
\begin{eqnarray}
  M_{\rm miss}^2 &=& (p_{e^+ e^-} - p_{\rm tag} - p_{D^*} - p_{\ell})^2,
\end{eqnarray} 
where $p_\ell$ denotes the four-momentum of the charged lepton and the other variables were defined earlier. The normalization events populate the region near $M_{\rm miss}^2 = 0~{\rm GeV}^2$ because there is exactly one neutrino in an event. We require $-0.5 < M_{\rm miss}^2 < 0.5~{\rm GeV}^2$. We further require $E_{\rm ECL}$ to be less than 1.5~GeV.

Finally, for both the signal and the normalization events, we require that there be no extra charged tracks with $dr < 5~{\rm cm}$ and $|dz| < 20~{\rm cm}$, and normal $\pi^0$ candidates.

\subsection{Best candidate selection}

After event reconstruction, the average number of retained candidates per event is about 1.09 for charged $B$ mesons and 1.03 for neutral $B$ mesons. In events where two or more candidates are reconstructed, 2.1 candidates are found on average. Multiple-candidate events mostly arise from more than one combination of a $D$ candidate with photons or soft pions. For the charged $B$ mode, about 2\% of the events are reconstructed both in the $D^{*0} \rightarrow D^0 \gamma$ and $D^{*0} \rightarrow D^0 \pi^0$ modes. Since the latter mode has a much higher branching fraction, we assign these events to the $D^{*0} \rightarrow D^0 \pi^0$ sample. The contribution of this type of multiple-candidate events is negligibly small in the neutral $B$ mode. We then select the most signal-like candidate as follows. For the $D^{*0} \rightarrow D^0 \gamma$ events, we select the candidate with the most energetic photon associated with the $D^0$. For the $D^{*0} \rightarrow D^0 \pi^0$ and $D^{*+} \rightarrow D^+ \pi^0$ events, we select the candidate with the soft $\pi^0$ that has an invariant mass nearest the nominal $\pi^0$ mass. For the $D^{*+} \rightarrow D^0 \pi^+$ events, we select one candidate at random since the multiple-candidate probability is only $\mathcal{O}(0.01\%)$. After the $D^*$ candidate selection, roughly 2\% of the retained events are reconstructed both in the $\tau^- \rightarrow \pi^- \nu_\tau$ and the $\tau^- \rightarrow \rho^- \nu_{\tau}$ samples. Since the MC study indicates that about 80\% of such events originate from the $\tau^- \rightarrow \rho^- \nu_\tau$ decay, we assign these events to the $\tau^- \rightarrow \rho^- \nu_\tau$ sample.

\subsection{Sample composition}\label{sec:sample}

The reconstructed events are categorized in turn as below. Based on this categorization, we construct histogram probability density functions (PDFs) from the MC samples to perform a final fit.
\begin{description}
\item[Signal]\mbox{}\\
  Correctly reconstructed signal events that originate from $\tau^- \rightarrow \pi^- (\rho^-) \nu_\tau$ events are categorized in this component. The yield is treated as a free parameter determined by $R(D^*)$ and $P_{\tau}(D^*)$.
\item[\textbf{\boldmath$\rho \leftrightarrow \pi$} cross feed]\mbox{}\\
  Cross feed events, where the $\tau^- \rightarrow \rho^- \nu_\tau$ events are reconstructed as $\tau^- \rightarrow \pi^- \nu_\tau$ due to misreconstruction of one $\pi^0$, or the $\tau^- \rightarrow \pi^- \nu_\tau$ events are reconstructed as $\tau^- \rightarrow \rho^- \nu_\tau$ by adding a random $\pi^0$, comprise this component. Since these events originate from ${\bar B} \rightarrow D^* \tau^- \bar{\nu}_\tau$, they contribute to the $R(D^*)$ determination. They are also used for the $P_\tau(D^*)$ determination after the bias on $P_\tau(D^*)$ is corrected by MC information.
\item[Other \textbf{\boldmath$\tau$} cross feed]\mbox{}\\
  $\bar{B} \rightarrow D^* \tau^- \bar{\nu}_\tau$ events with other $\tau$ decay modes also contribute to the signal sample. They originate mainly from $\tau^- \rightarrow a_1^- (\rightarrow \pi^- \pi^0 \pi^0) \nu_\tau$ with one or two missing $\pi^0$, or $\tau^- \rightarrow \mu^- \bar{\nu}_\mu \nu_\tau$ with a low-momentum $\mu^-$ that does not reach the KLM. These two modes occupy about 80\% of this component. The MC study shows that the cross feed events both from $\tau^- \rightarrow a_1^- \nu_\tau$ and $\tau^- \rightarrow \mu^- \bar{\nu}_\mu \nu_\tau$ have negligible impact on our $P_\tau(D^*)$ measurement. In the fit, the yield of this category is determined by $R(D^*)$.
\item[\textbf{\boldmath$\bar{B} \rightarrow D^* \ell^- \bar{\nu}_\ell$}]\mbox{}\\
  The decay $\bar{B} \rightarrow D^* \ell^- \bar{\nu}_\ell$ contaminates the signal sample due to misassignment of $\ell^-$ as $\pi^-$. We fix the $\bar{B} \rightarrow D^* \ell^- \bar{\nu}_\ell$ yield in the signal sample from the fit to the $M_{\rm miss}^2$ distribution of the normalization sample.
\item[\textbf{\boldmath$\bar{B} \rightarrow D^{**} \ell^- \bar{\nu}_\ell$} and hadronic \textbf{\boldmath$B$} decays]\mbox{}\\
  The $\bar{B} \rightarrow D^{**} \ell^- \bar{\nu}_\ell$ ($\bar{B} \rightarrow D^{**} \tau^- \bar{\nu}_\tau$ is also included in this category) and hadronic $B$ decays are the most uncertain component due to limited experimental knowledge. By missing a few particles such as $\pi^0$ mesons, the event topology resembles the signal event. We combine these decay modes into one component. The fractions of the $\bar{B} \rightarrow D^{**} \ell^- \bar{\nu}_\ell$ decays and hadronic $B$ decays are about 10\% and 90\%, respectively, according to the MC study. Since it is difficult to estimate the yield of this component using MC simulation or to fix the yield using control data samples, we float the yield in the final fit. One exception is the collection of modes with two charm mesons such as $\bar{B} \rightarrow D^* D_s^{(*)-}$ and $\bar{B} \rightarrow D^* \bar{D}^{(*)} K^-$. Since the branching fractions of these modes have been studied experimentally, we fix their yield using the MC expectation after correction with the branching fractions based on Ref.~\cite{cite:PDG:2016}.
\item[Continuum]\mbox{}\\
  Continuum events from the $e^+ e^- \rightarrow q{\bar q}$ process provide a minor contribution at $\mathcal{O}(0.1\%)$ in the signal sample. We fix the yield using the MC expectation.
\item[Fake \textbf{\boldmath$D^*$}]\mbox{}\\
  All events containing fake $D^*$ candidates are categorized in this component. This is the main background source in the charged $B$ meson sample. For the neutral $B$ sample, many $D^{*+}$ candidates are reconstructed from the combination of a $D^0$ with a $\pi^\pm$ and therefore much more cleanly reconstructed than the other $D^*$ modes with $\pi^0$ or $\gamma$. The yield is determined from a comparison of the data and the MC sample in the $\Delta M$ sideband regions.
\end{description}

\subsection{Measurement Method of \textbf{\boldmath$R(D^*)$} and \textbf{\boldmath$P_\tau(D^*)$}}\label{subsec:sig-recon}

We use the following variables to measure yields of the signal and the normalization modes. For the normalization mode, $M_{\rm miss}^2$ is the most suitable variable due to its high purity. On the other hand, the shape of the $M_{\rm miss}^2$ distribution for the signal mode has a strong correlation with $P_\tau(D^*)$. To measure the signal yield, we use $E_{\rm ECL}$ because it has a small correlation to $P_\tau(D^*)$ and provides good discrimination between the signal and the background modes.

The value of $R(D^*)$ is measured using the formula
\begin{eqnarray}
  R(D^*) &=& \frac{\epsilon_{\rm norm}^j N_{\rm sig}^{ij}} {\mathcal{B}_\tau^i \epsilon_{\rm sig}^{ij} N_{\rm norm}^j},\label{eq:Rst}
\end{eqnarray}
where $\mathcal{B}_\tau^i$ denotes the relevant $\tau$ branching fraction, and $\epsilon_{\rm sig}^{ij}$ and $\epsilon_{\rm norm}^j$ ($N_{\rm sig}^{ij}$ and $N_{\rm norm}^j$) are the efficiencies (the observed yields) for the signal and the normalization modes, respectively. The indices $i$ and $j$ represent the $\tau$ decays ($\tau^- \rightarrow \pi^- \nu_\tau$ or $\rho^- \nu_\tau$) and the $B$ charges (charged $B$ or neutral $B$), respectively. Assuming isospin symmetry, we use $R(D^*) = R(D^{*0}) = R(D^{*+})$.

\begin{figure*}[t!]
  \centering
  \includegraphics[width=7.5cm]{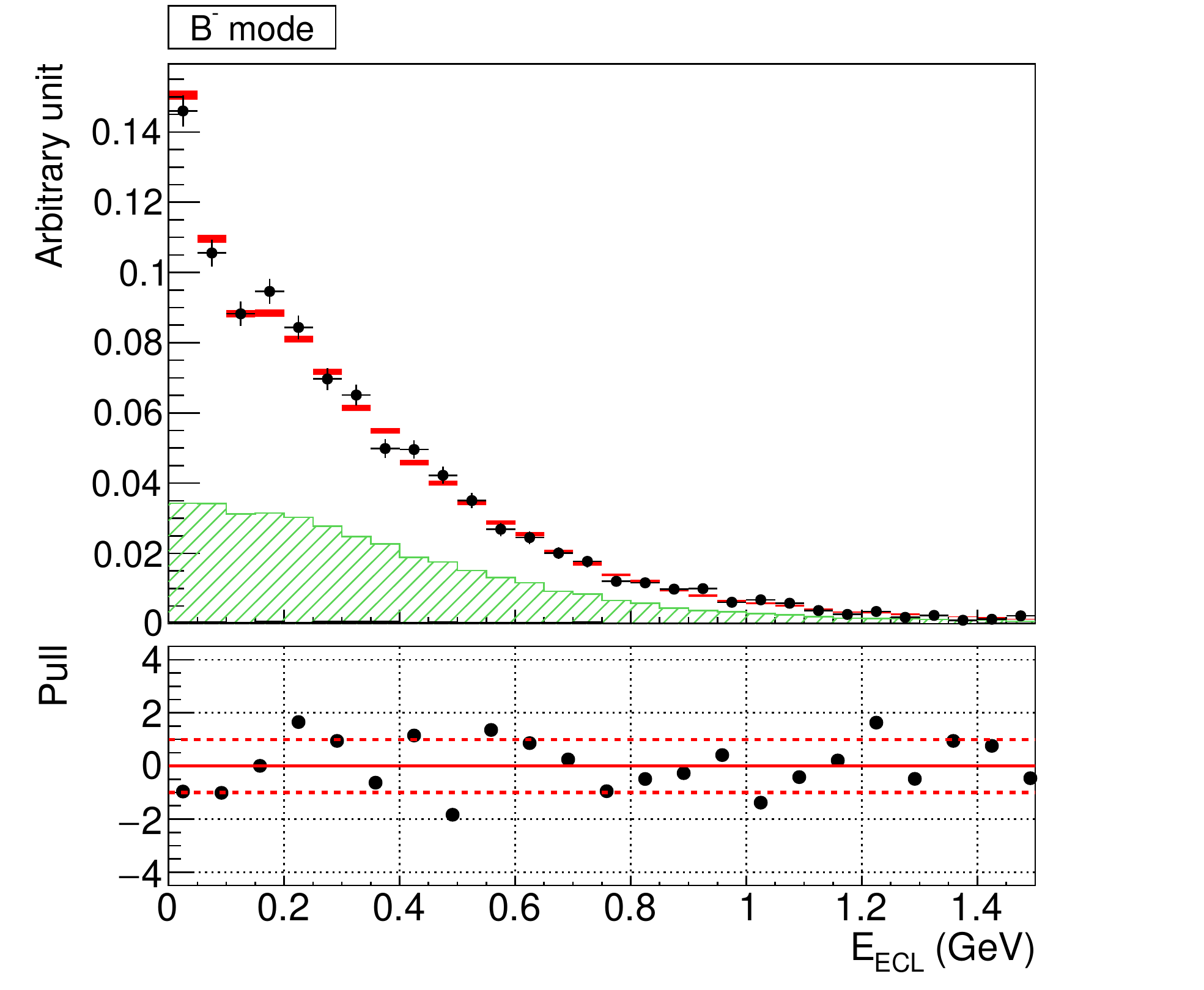}
  \includegraphics[width=7.5cm]{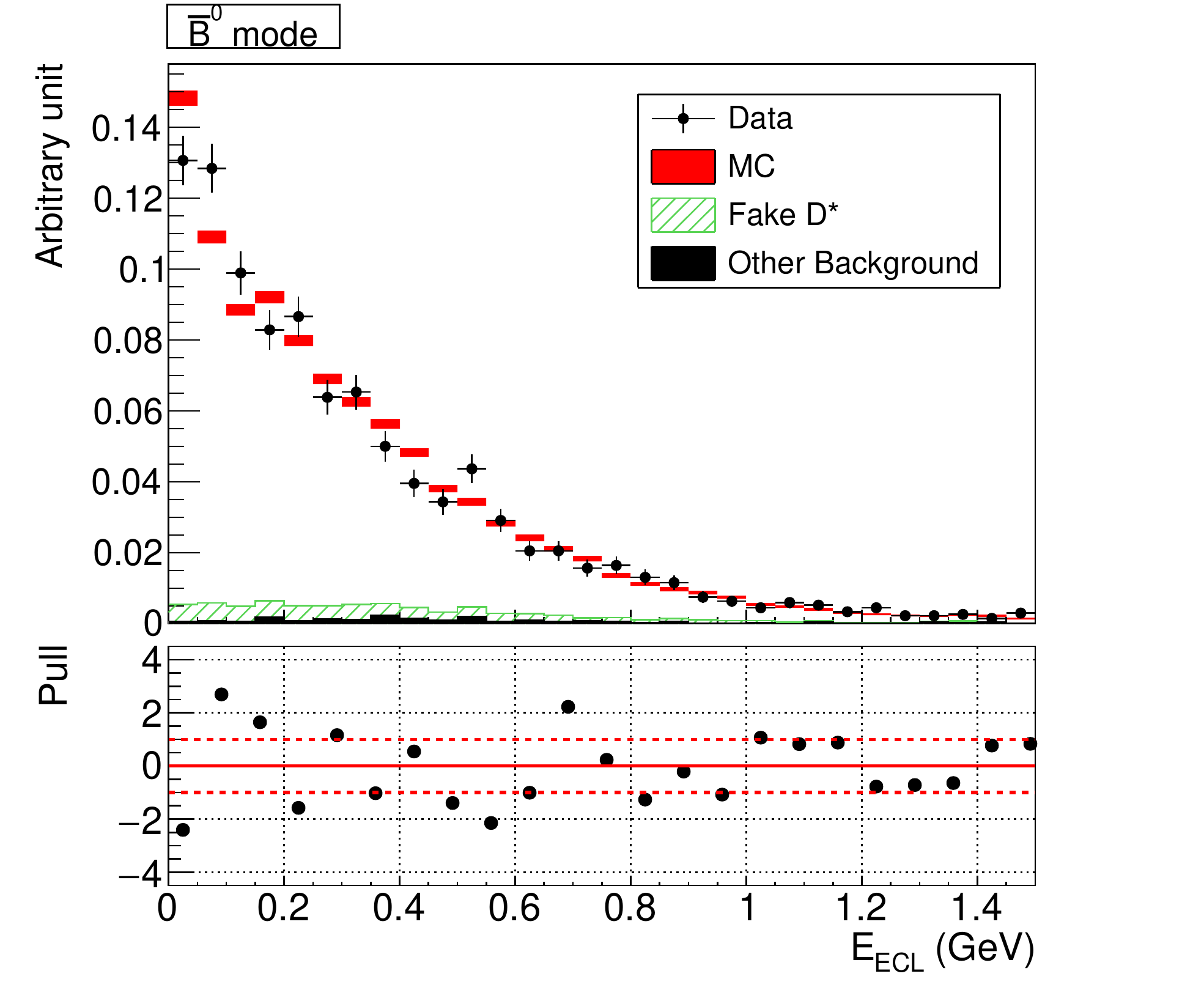}
  \caption{Comparison of the $E_{\rm ECL}$ distributions between the data (black circles) and the MC simulation (red rectangles) of the normalization mode. The area of the histograms are normalized to unity.}
  \label{fig:EECL_Dstlnu_comp}
\end{figure*}

The value of $P_\tau(D^*)$ is determined using the formula
\begin{eqnarray}
  P_\tau(D^*) &=& \frac{2}{\alpha_i} \frac{N_{\rm sig}^{Fij} - N_{\rm sig}^{Bij}} {N_{\rm sig}^{Fij} + N_{\rm sig}^{Bij}},\label{eq:Pst}
\end{eqnarray}
where $N_{\rm sig}^{F(B)ij}$ denotes the signal yield in the region $\cos\theta_{\rm hel} > (<)~0$ and satisfies $N_{\rm sig}^{Fij} + N_{\rm sig}^{Bij} = N_{\rm sig}^{ij}$. This formula is obtained by calculating
\begin{eqnarray}
  N_{\rm sig}^{Fij} &=& N_{\rm sig}^{ij} \int_{0}^{1} \frac{d \Gamma^{ij}(D^{(*)})}{d \cos\theta_{\rm hel}} d \cos\theta_{\rm hel},\\
  N_{\rm sig}^{Bij} &=& N_{\rm sig}^{ij} \int_{-1}^{0} \frac{d \Gamma^{ij}(D^{(*)})}{d \cos\theta_{\rm hel}} d \cos\theta_{\rm hel}.
\end{eqnarray}
The differential decay rate $d \Gamma^{ij}(D^{(*)}) / d \cos\theta_{\rm hel}$ is given by Eq.~(\ref{eq:coshel}). As with $R(D^*)$, we use the common parameters $P_\tau(D^*) = P_\tau(D^{*0}) = P_\tau(D^{*+})$.

Due to detector efficiency effects, the measured polarization, $P_\tau^{\rm raw}(D^*)$, is biased from the true value of $P_\tau(D^*)$. To correct for this bias, we form a linear function that maps $P_\tau(D^*)$ to $P_\tau^{\rm raw}(D^*)$ using several MC sets with different $P_\tau(D^*)$. This function, denoted the $P_\tau(D^*)$ correction function, is separately prepared for each $\tau$ sample since the detector bias depends on the given $\tau$ mode. We also make a $P_\tau(D^*)$ correction function for the $\rho \leftrightarrow \pi$ cross feed component  to take into account the distortion of the $\cos\theta_{\rm hel}$ distribution shape. In the $P_\tau(D^*)$ correction, other kinematic distributions are assumed to be consistent with the SM predictions.

\begin{figure*}[t!]
  \centering
  \subfigure[]{\includegraphics[width=5.5cm]{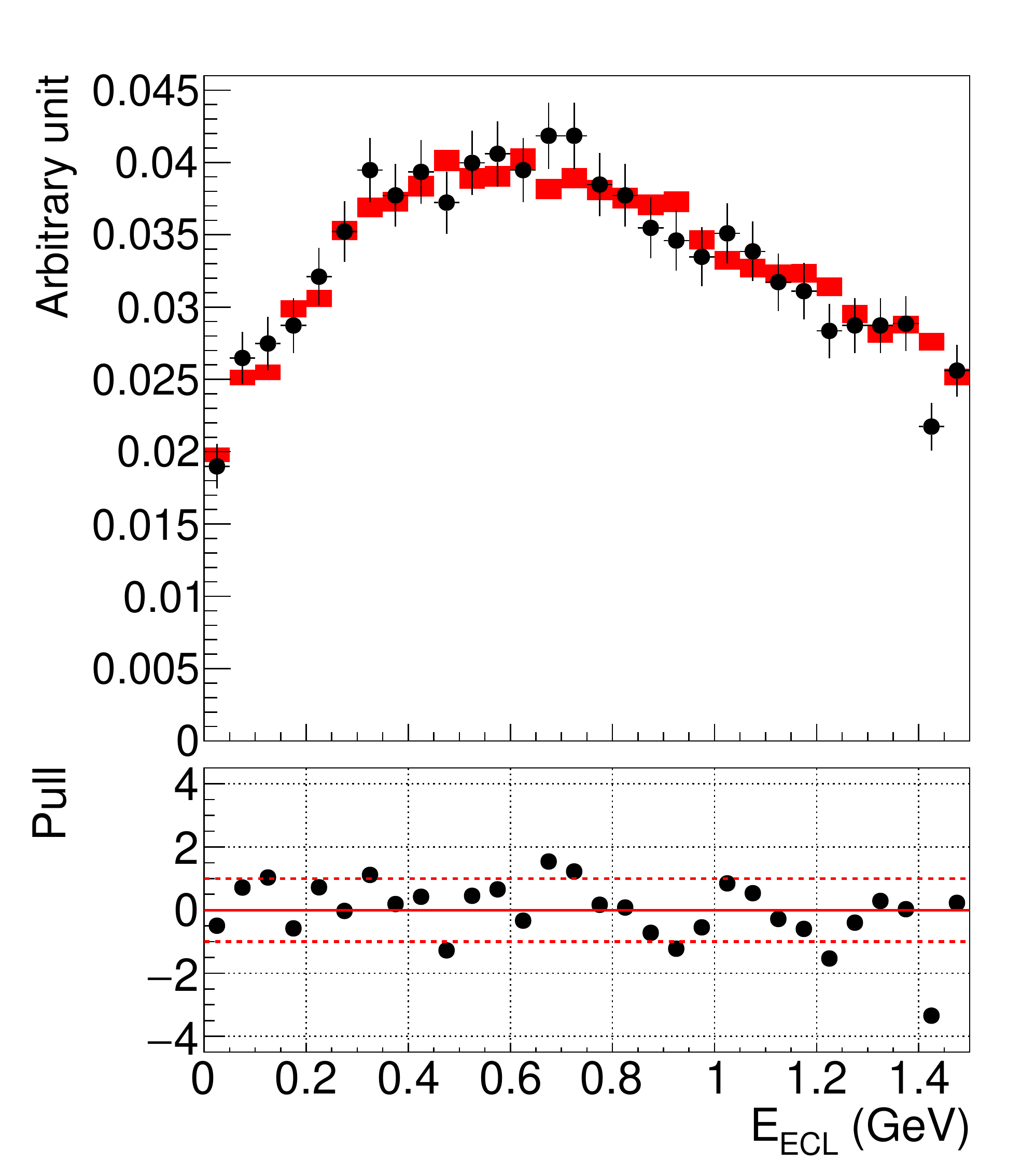}\label{fig:FakeDst-EECL-comp}}
  \subfigure[]{\includegraphics[width=5.5cm]{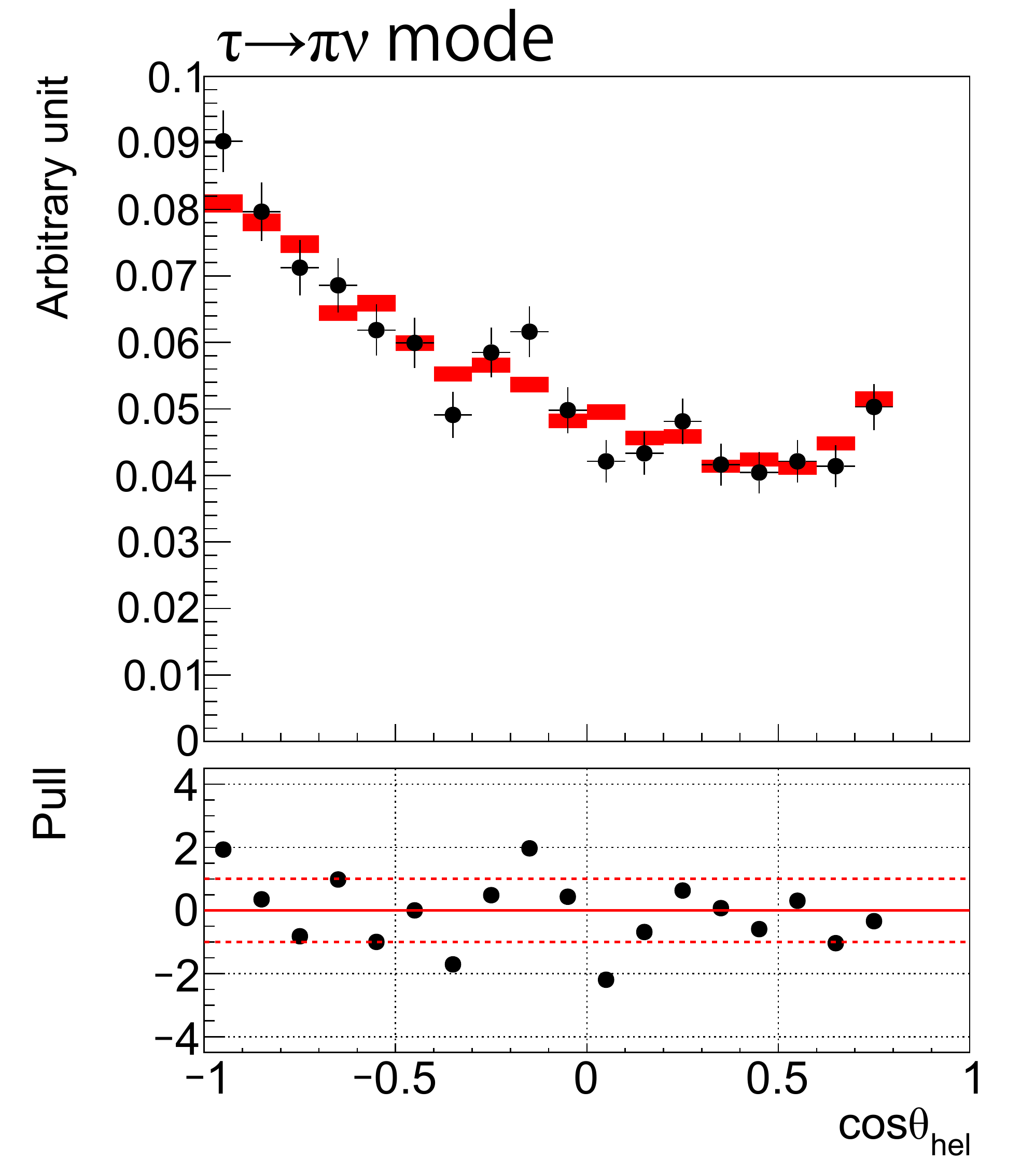}\label{fig:FakeDst-coshel-pinu-comp}}
  \subfigure[]{\includegraphics[width=5.5cm]{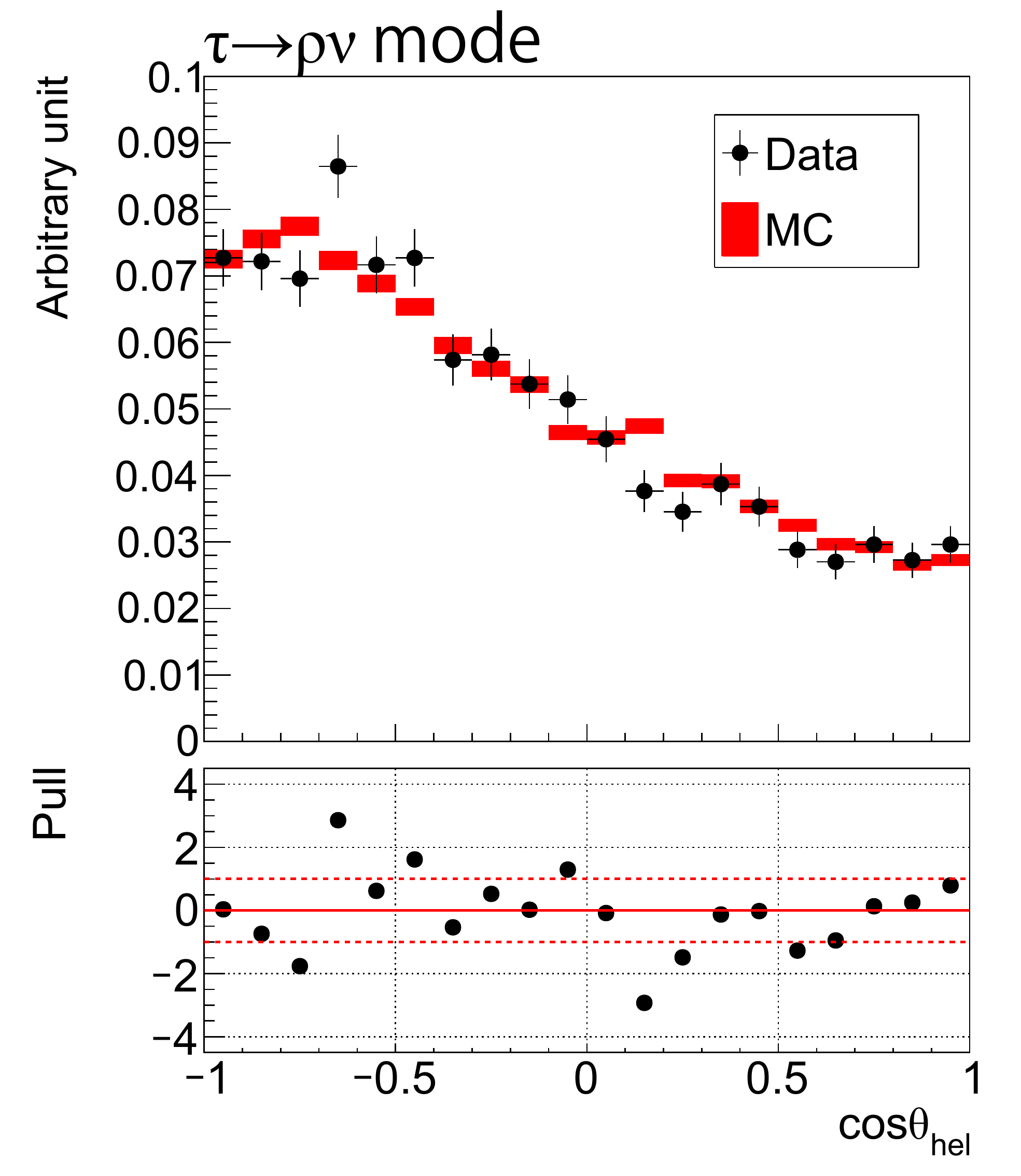}\label{fig:FakeDst-coshel-rhonu-comp}}
  \caption{Comparisons between the data (black circles) and the MC simulation (red rectangles) in the $\Delta M$ sideband regions, where the distributions are normalized to unity. (a) $E_{\rm ECL}$ distribution, (b) $\cos\theta_{\rm hel}$ distribution for the $\tau^- \rightarrow \pi^- \nu_\tau$ mode, (c) $\cos\theta_{\rm hel}$ distribution for the $\tau^- \rightarrow \rho^- \nu_\tau$ mode. All the corresponding channels are combined.}
  \label{fig:FakeDst-comp}
  \vspace{5mm}
\end{figure*}

\section{Background Calibration and PDF Validation}

To use the MC distributions as histogram PDFs, the MC simulation needs to be verified using calibration data samples. In this section, the calibration of the PDF shapes is discussed.

\subsection{Signal PDF shape}

To validate the $E_{\rm ECL}$ shape of the signal component, we use the normalization mode as the control sample. It has similar $E_{\rm ECL}$ properties to the signal component; there is no extra photon from the $B_{\rm sig}$ decay except for bremsstrahlung photons, and therefore the $E_{\rm ECL}$ shape is mostly determined by the background photons. The normalization sample contains about 50 times more events than the expected signal yield. Figure~\ref{fig:EECL_Dstlnu_comp} shows a comparison of $E_{\rm ECL}$ between data and MC simulation. The pull of each bin is shown in the bottom panel; hereinafter, the pull in the $i$th bin is defined as
\begin{eqnarray}
  \text{Pull}^i &=& \frac{N_{\rm data}^i - N_{MC}^i}{\sqrt{(\sigma_{\rm data}^i)^2 + (\sigma_{\rm MC}^i)^2}},
\end{eqnarray}
where $N_{\rm data (MC)}^i$ and $\sigma_{\rm data (MC)}^i$ denote the number of events and the statistical error, respectively, in the $i$th bin of the data (MC) distribution. The fake $D^*$ yield is scaled based on the calibration discussed in the next section. Since the contribution from the other background components is negligibly small, it is fixed to the MC expectation. The ${\rm ECL}$ shape in the MC sample agrees well with the data within statistical uncertainty.

\subsection{Fake \textbf{\boldmath$D^*$} events}\label{sec:FakeDst}

One of the most significant background components arises from fake $D^*$ candidates. The combinatorial fake $D^*$ background processes are difficult to model precisely in the MC simulation. The $E_{\rm ECL}$ shapes for the data and the MC sample are compared using $\Delta M$ sideband regions of 50--500~MeV, 135--190~MeV, 135--190~MeV, and 140--500~MeV for $D^{*0} \rightarrow D^0 \gamma$, $D^{*0} \rightarrow D^0 \pi^0$, $D^{*+} \rightarrow D^+ \pi^0$, and $D^{*+} \rightarrow D^0 \pi^+$, respectively; each excludes about $\pm 4\sigma$ around the $\Delta M$ peak. These sideband regions contain 5 to 50 times more events than the signal region. Figure~\ref{fig:FakeDst-EECL-comp} shows the comparison of the $E_{\rm ECL}$ shapes. Although all the $D^*$ and $\tau$ modes are combined in these figures, the $E_{\rm ECL}$ shape has been compared in 16 subsamples of $B$ modes, $D^*$ modes, $\tau$ modes, and the two $\cos\theta_{\rm hel}$ regions. We find good agreement of the $E_{\rm ECL}$ shape within the statistical uncertainty of these mass sideband data samples. We also check the $\cos\theta_{\rm hel}$ distribution in the $\Delta M$ sideband region, as shown in Figs.~\ref{fig:FakeDst-coshel-pinu-comp} and \ref{fig:FakeDst-coshel-rhonu-comp}. The $\cos\theta_{\rm hel}$ distribution in the MC simulation also shows good agreement with the data within the statistical uncertainty.

In both the signal and the normalization samples, yield discrepancies of up to 20\% are observed. The fake $D^*$ yields in the signal region of the MC simulation are scaled by the yield ratios of the data to the MC sample in the $\Delta M$ sideband regions.

\subsection{\textbf{\boldmath$\bar{B} \rightarrow D^{**} \ell^- \bar{\nu}_\ell$} and hadronic \textbf{\boldmath$B$} composition}\label{sec:background}

\begin{figure}[t!]
  \centering
  \includegraphics[width=9cm]{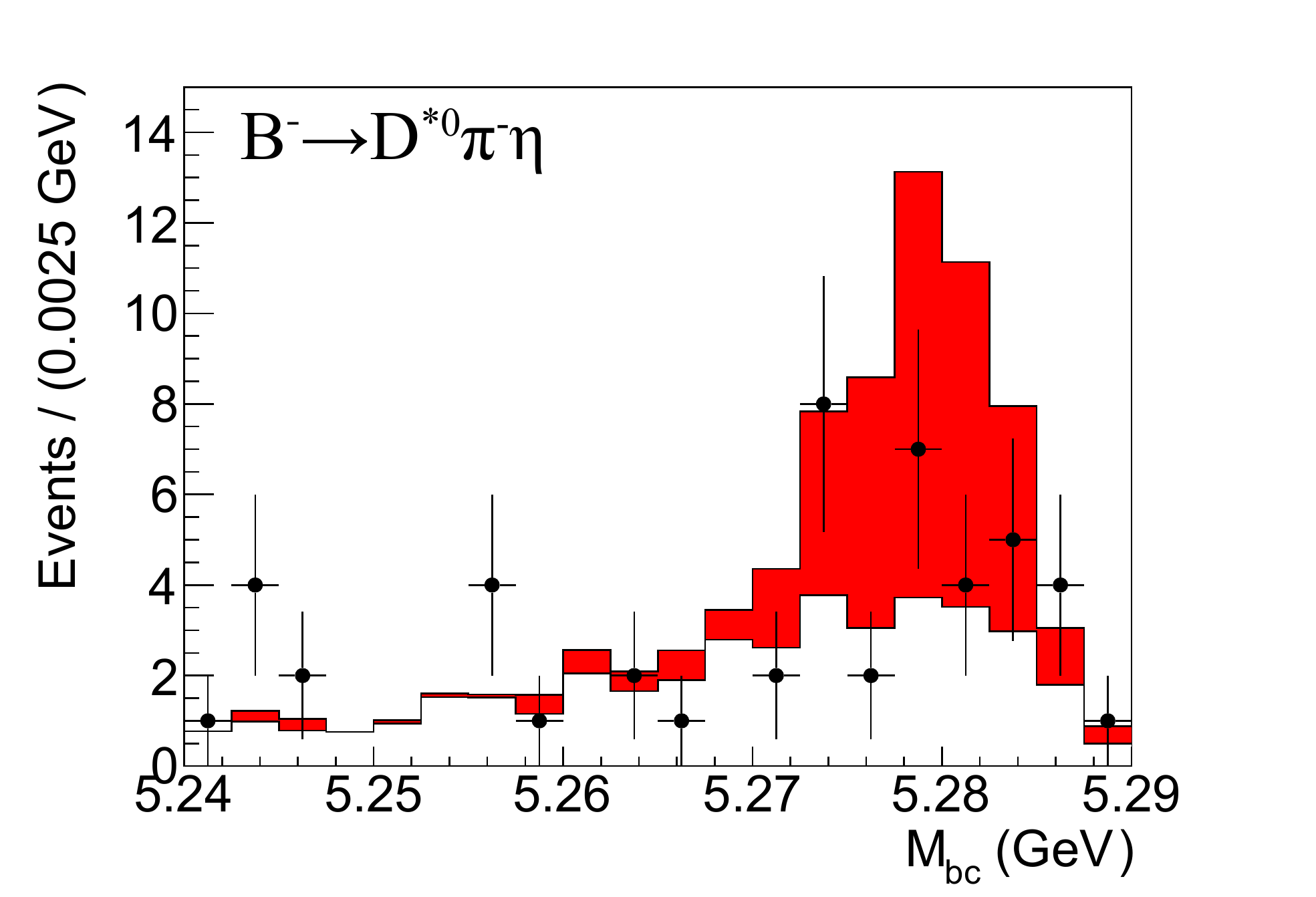}
  \caption{Distribution of the $M_{\rm bc}^{\rm sig}$ for the $B^- \rightarrow D^{*0} \pi^- \eta$ sample. The solid histogram shows the MC distribution (the red-filled and the white components for the correct $B$ and fake $B$ candidates, respectively) and the black dots are the data distribution.}
  \label{fig:calib-Mbc}
\end{figure}

\begin{table}[t!]
  \centering
  \caption{Calibration factors used to correct the hadronic $B$ background rates in the MC simulation. The errors arise from the calibration sample statistics.}
  \vspace{3mm}
  \begin{tabular}{@{\hspace{0.5cm}}l@{\hspace{0.5cm}}|@{\hspace{0.5cm}}c@{\hspace{0.5cm}}@{\hspace{0.5cm}}c@{\hspace{0.5cm}}}
    \hline
    \hline
    $B$ decay mode & $B^-$ & ${\bar B^0}$\\
    \hline
    $D^* \pi^- \pi^- \pi^+$             & $< 0.51$                 & $0.62 ^{+0.67} _{-0.49}$\\ 
    $D^* \pi^- \pi^- \pi^+ \pi^0$       & $0.31 ^{+0.43} _{-0.40}$ & $0.59 ^{+0.45} _{-0.39}$\\ 
    $D^* \pi^- \pi^- \pi^+ \pi^0 \pi^0$ & $2.15 ^{+1.70} _{-1.60}$ & $2.60 ^{+6.95} _{-2.24}$\\ 
    $D^* \pi^- \pi^0$                   & $0.06 ^{+0.33} _{-0.28}$ & $<0.47$\\ 
    $D^* \pi^- \pi^0 \pi^0$             & $0.09 ^{+1.04} _{-0.98}$ & $1.63 ^{+0.74} _{-0.69}$\\ 
    $D^* \pi^- \eta$                    & $0.24 ^{+0.21} _{-0.18}$ & $0.15 ^{+0.16} _{-0.10}$\\
    $D^* \pi^- \eta \pi^0$              & $0.74 ^{+0.79} _{-0.75}$ & $0.89 ^{+1.04} _{-0.88}$\\ 
    \hline
    \hline
  \end{tabular}
  \label{tab:calib}
\end{table}

As discussed in Sec.~\ref{sec:sample}, the yield of the $\bar{B} \rightarrow D^{**} \ell^- \bar{\nu}_\ell$ and hadronic $B$ background component is determined in the final fit. The PDF shape of this background must be corrected with data, as a change in the $B$ decay composition may modify the $E_{\rm ECL}$ shape and thereby introduce bias in the measurements of $R(D^*)$ and $P_\tau(D^*)$.

If a background $B$ decay contains a $K_L^0$ in the final state, it may peak in the $E_{\rm ECL}$ signal region. We correct the branching fractions of the ${\bar B} \rightarrow D^* \pi^- K_L^0$ and ${\bar B} \rightarrow D^* K^- K_L^0$ modes in the MC simulation using the measured values~\cite{cite:PDG:2016,cite:Belle_DstKKL:2002}. We do not apply branching fraction corrections for the other decays with $K_L^0$ because they have relatively small expected yields. However, we assume 100\% of the uncertainty on the branching fractions to estimate systematic uncertainties, as discussed in Sec.~\ref{sec:syst}.

\begin{figure}[t!]
  \centering
  \subfigure[]{\includegraphics[width=7.5cm]{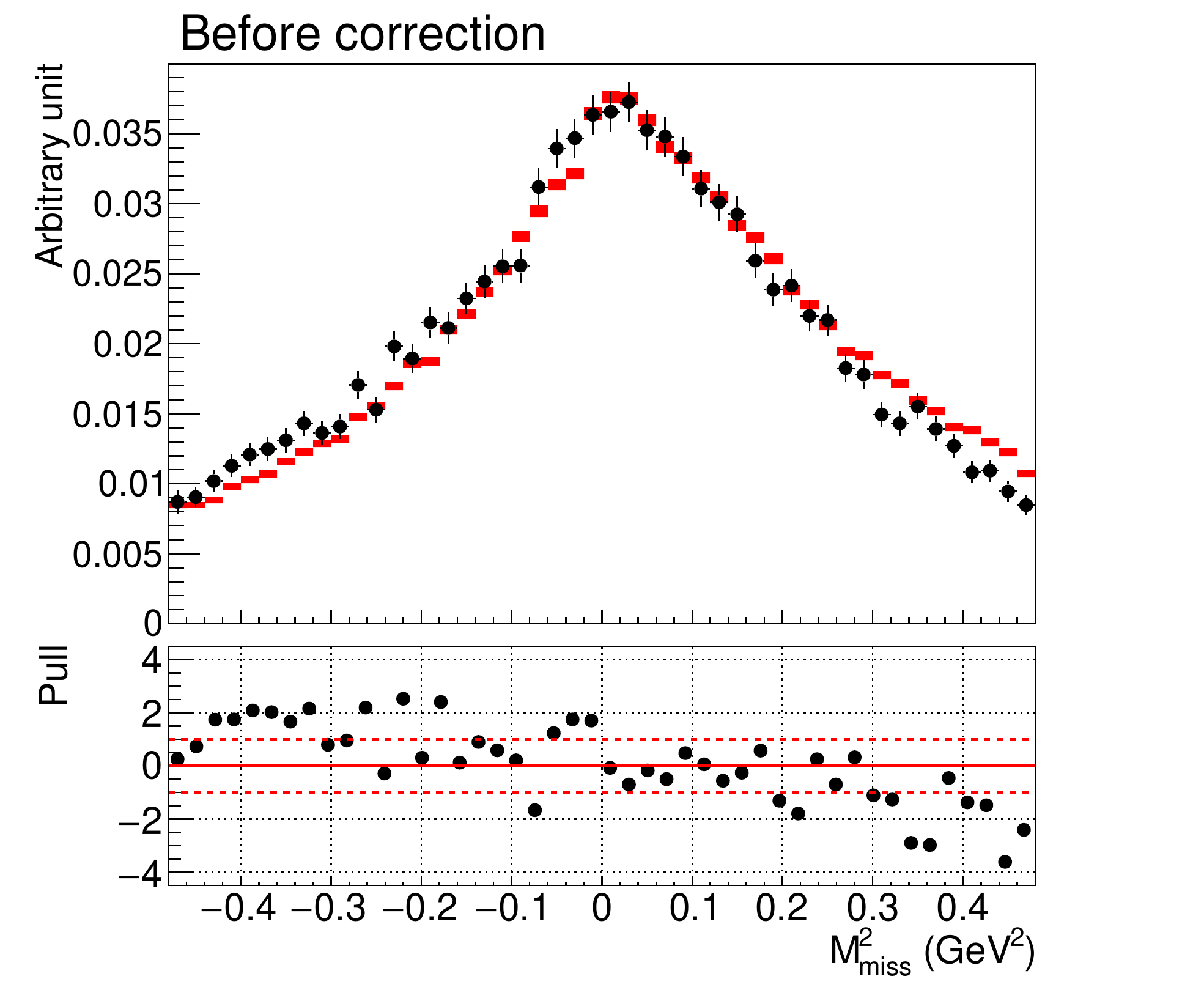}\label{fig:FakeDst-Mmiss2-comp-before}}
  \subfigure[]{\includegraphics[width=7.5cm]{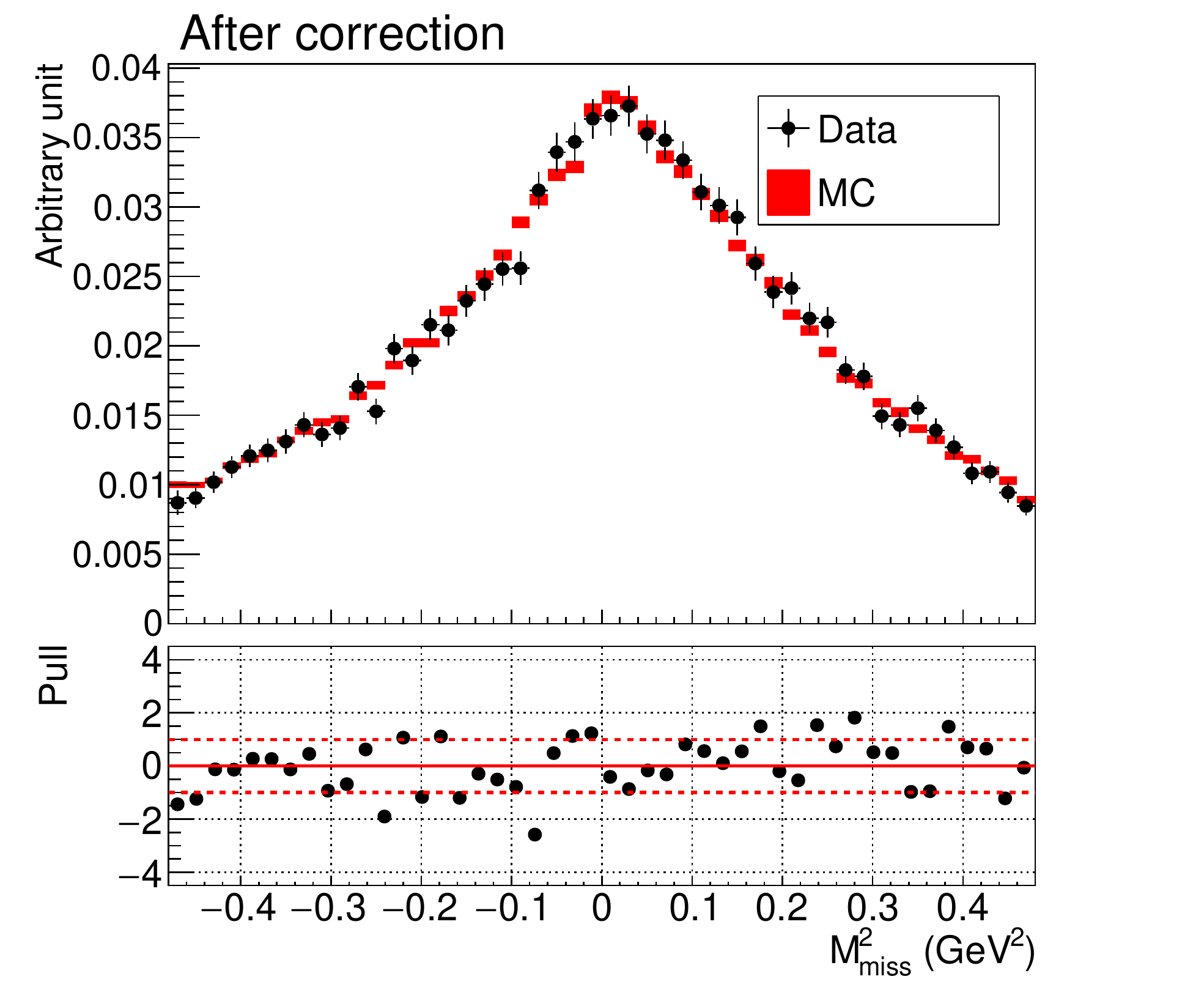}\label{fig:FakeDst-Mmiss2-comp-after}}
  \caption{Comparison of the $M_{\rm miss}^2$ distributions between the data (black circles) and the MC simulation (red rectangles) in the $\Delta M$ sideband regions of the $D^{*0}$ channels: (a) before the shape correction, and (b) after the correction. All the distributions are normalized to unity.}
  \label{fig:FakeDst-Mmiss2-comp}
\end{figure}

\begin{figure*}[t!]
  \centering
  \includegraphics[width=18cm]{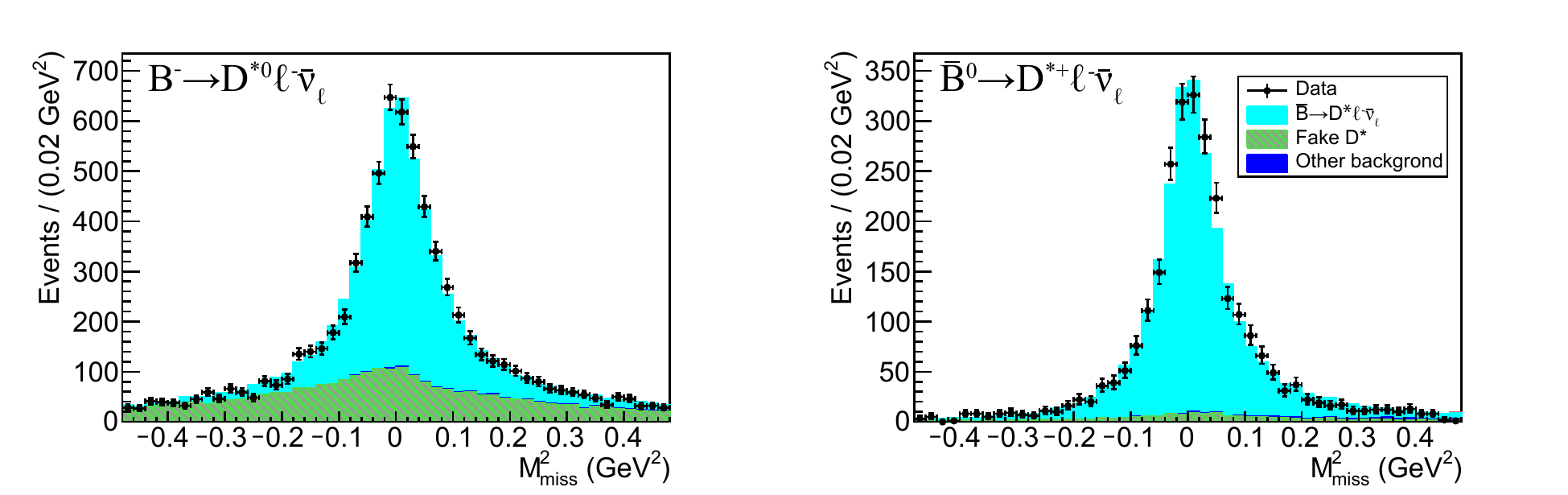}
  \caption{Fit result to the normalization samples.}
  \label{fig:calib-normalization-fit}
\end{figure*}

Other types of hadronic $B$ decay background often contain neutral particles such as $\pi^0$ or $\eta$ as well as pairs of charged particles. We calibrate the rate of hadronic $B$ decays in the signal region based on control samples where one $B$ is fully reconstructed with the hadronic tag, and the signal side is reconstructed in seven final states (${\bar B} \rightarrow D^* \pi^- \pi^- \pi^+$, ${\bar B} \rightarrow D^* \pi^- \pi^- \pi^+ \pi^0$, ${\bar B} \rightarrow D^* \pi^- \pi^- \pi^+ \pi^0 \pi^0$, ${\bar B} \rightarrow D^* \pi^- \pi^0$, ${\bar B} \rightarrow D^* \pi^- \pi^0 \pi^0$, ${\bar B} \rightarrow D^* \pi^- \eta$, and ${\bar B} \rightarrow D^* \pi^- \eta \pi^0$). Charged and neutral $B$ mesons are reconstructed separately. Pairs of photons with an invariant mass ranging from 500 to 600~MeV are selected as $\eta$ candidates. We then extract the yield of the data and the MC sample in the region $q^2 > 4~{\rm GeV}^2$ and $|\cos\theta_{\rm hel}| < 1$, which is the same requirement as in the signal sample. To calculate $\cos\theta_{\rm hel}$, we assume that (one of) the charged pion(s) is the $\tau$ daughter. The signal-side energy difference $\Delta E^{\rm sig}$ or the beam-energy-constrained mass $M_{\rm bc}^{\rm sig}$ of the $B_{\rm sig}$ candidate is used for the yield extraction. Figure~\ref{fig:calib-Mbc} shows the $M_{\rm bc}^{\rm sig}$ distribution for the $B^- \rightarrow D^* \pi^- \eta$ mode as an example. We estimate yield calibration factors by taking ratios of the yields in the data to that in the MC sample. If there is no observed signal event in the calibration sample, we assign a 68\% confidence level (C.L.) upper limit on the yield. The obtained calibration factors are summarized in Table~\ref{tab:calib}. Additionally, we correct the branching fractions of the decays $B^- \rightarrow D^{*+} \pi^- \pi^- \pi^0$, ${\bar B} \rightarrow D^* \omega \pi^-$ and ${\bar B} \rightarrow D^* {\bar p} n$ based on Refs.~\cite{cite:PDG:2016,cite:Belle_Dstomegapi:2015}.

About 80\% of the hadronic $B$ background is covered by the calibrations discussed above. We estimate the systematic uncertainties on our observables due to the uncertainties of the calibration factors in Sec.~\ref{sec:syst}.

\subsection{\textbf{\boldmath$M_{\rm miss}^2$} distribution for the normalization mode}\label{sec:Mmiss2}

In the fake $D^{*0}$ component of the charged $B$ channel, as shown in Fig.~\ref{fig:FakeDst-Mmiss2-comp-before}, we observe a slight discrepancy between the data and the MC sample. The $M_{\rm miss}^2$ discrepancy is therefore corrected based on this comparison. The $M_{\rm miss}^2$ distribution after the correction is shown in Fig.~\ref{fig:FakeDst-Mmiss2-comp-after}. The yield of the fake $D^*$ component is also corrected with the same method as applied to the signal sample.

After the correction for the fake $D^*$ component, we find that the $M_{\rm miss}^2$ resolution of the data sample is 10 to 20\% worse than that of the MC sample. We therefore smear the $M_{\rm miss}^2$ peak width to match that of the data sample. The correction is performed separately for each $D^*$ mode.

\section{Maximum Likelihood Fit}

An extended binned maximum likelihood fit is performed in two steps; we first perform a fit to the normalization sample to determine its yield, and then a simultaneous fit to eight signal samples from combinations of $(B^-, \bar{B}^0)$, $(\pi^- \nu_\tau, \rho^- \nu_\tau)$ and $(\cos\theta_{\rm hel} > 0, \cos\theta_{\rm hel} < 0)$. In the fit, $R(D^*)$ and $P_\tau(D^*)$ are common fit parameters among all the signal samples, while the $\bar{B} \rightarrow D^{**} \ell^- \bar{\nu}_\ell$ and hadronic $B$ yields are free to float.

Figure~\ref{fig:calib-normalization-fit} shows the fit result to the normalization sample. The $p$-value calculated from the agreement between the data and the fitted PDFs is 0.15. The normalization yields are measured to be $4711 \pm 81$ events for the charged $B$ sample and $2502 \pm 52$ events for the neutral $B$ sample, where the errors are statistical. As a cross check, we obtain the branching fractions of $(10.72 \pm 0.70)$\% for $B^- \rightarrow D^{*0} \ell^- \bar{\nu}_\ell$ and $(10.60 \pm 0.75)$\% for $\bar{B}^0 \rightarrow D^{*+} \ell^- \bar{\nu}_\ell$, where the values are the sum of $\bar{B} \rightarrow D^* e^- \bar{\nu}_e$ and $\bar{B} \rightarrow D^* \mu^- \bar{\nu}_\mu$. The error includes only a partial set of systematic uncertainties. These are consistent with the world averages $\mathcal{B}(B^- \rightarrow D^{*0} \ell^- \bar{\nu}_\ell) = (11.18 \pm 0.04 \pm 0.38)\%$ and $\mathcal{B}(\bar{B}^0 \rightarrow D^{*+} \ell^- \bar{\nu}_\ell) = (9.75 \pm 0.02 \pm 0.20)\%$, respectively~\cite{cite:HFLAV:2016}.

\begin{figure*}[p!]
  \centering
  \includegraphics[angle=90,width=6cm]{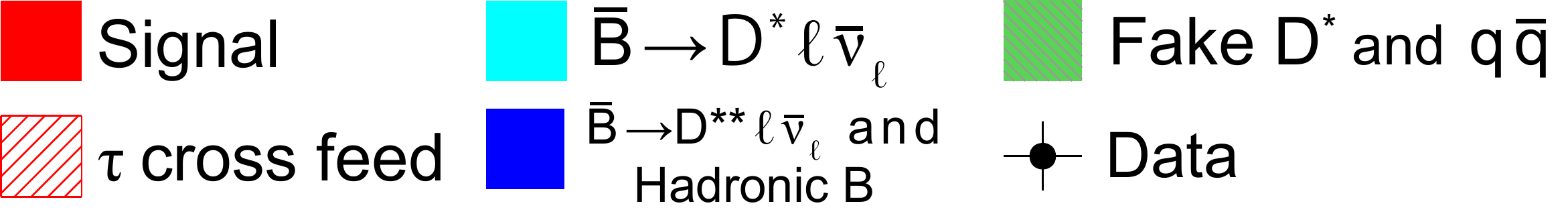}\\
  \includegraphics[width=15.5cm]{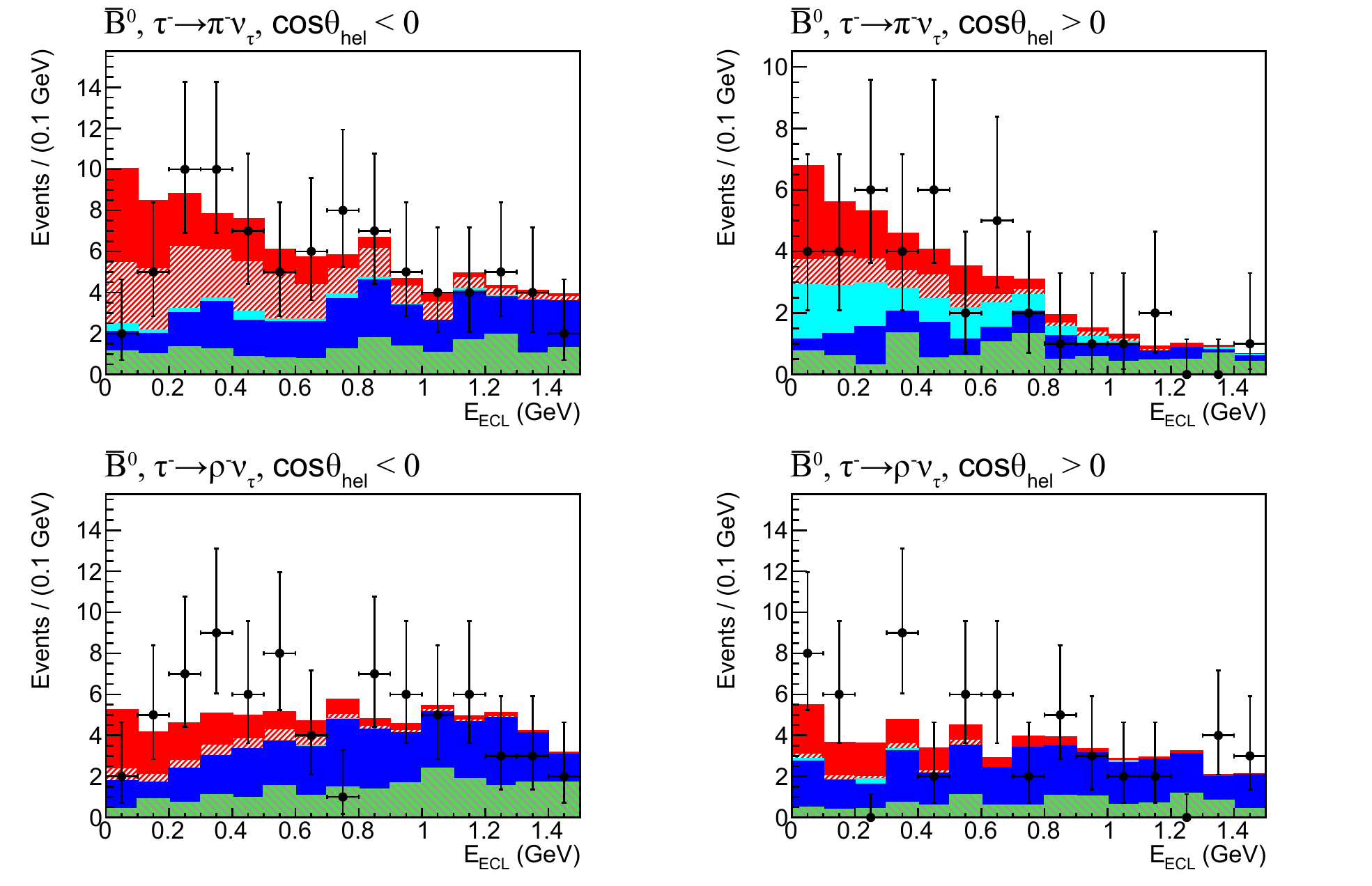}
  \includegraphics[width=15.5cm]{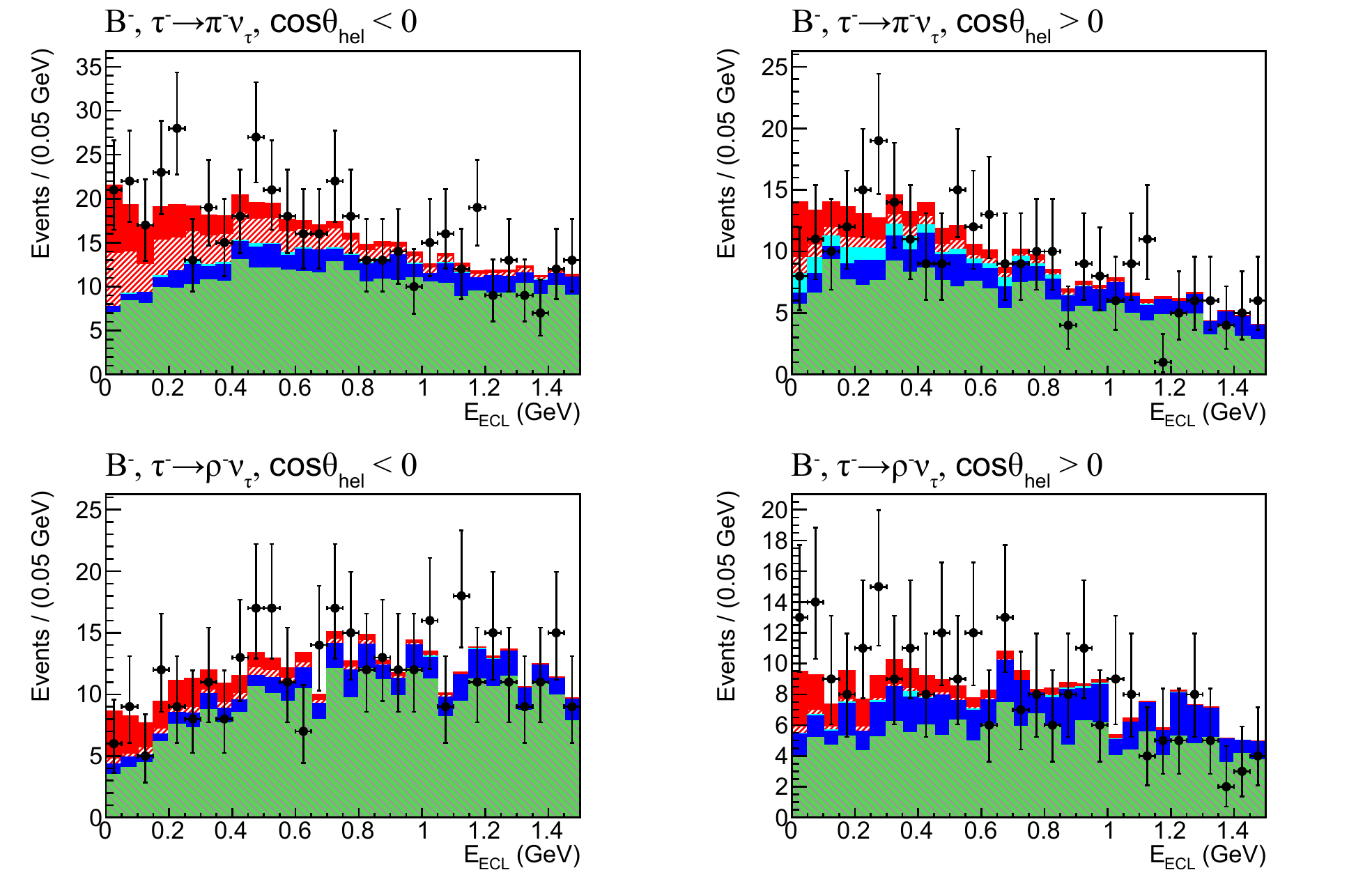}
  \caption{Fit results to the signal samples. The red-hatched ``$\tau$ cross feed'' combines the $\rho \leftrightarrow \pi$ cross feed and the other $\tau$ cross feed components.}
  \label{fig:fitresult}
\end{figure*}

\begin{figure*}[t!]
  \centering
  \includegraphics[angle=90,width=6cm]{Label.pdf}\\
  \includegraphics[width=7.5cm]{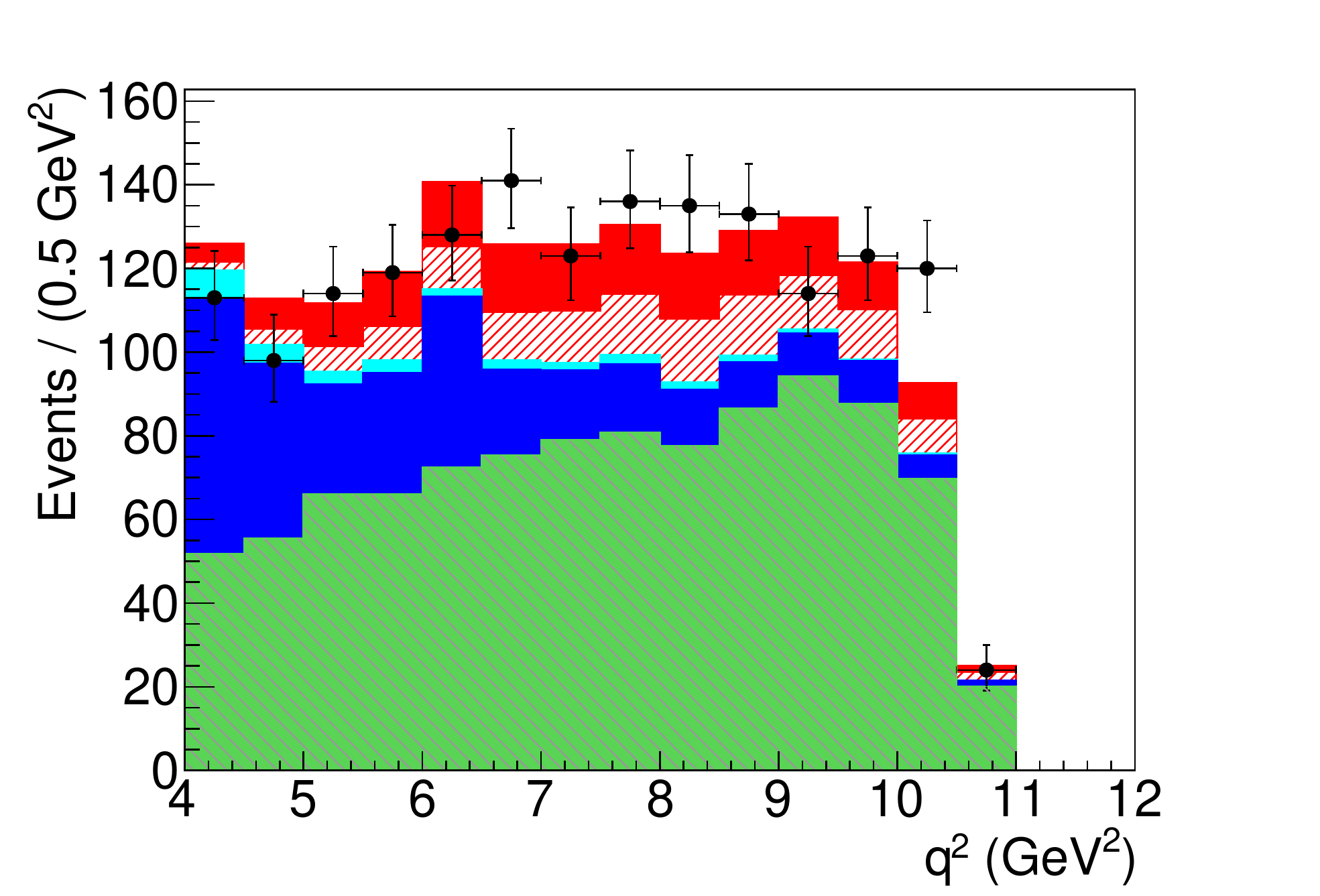}
  \includegraphics[width=7.5cm]{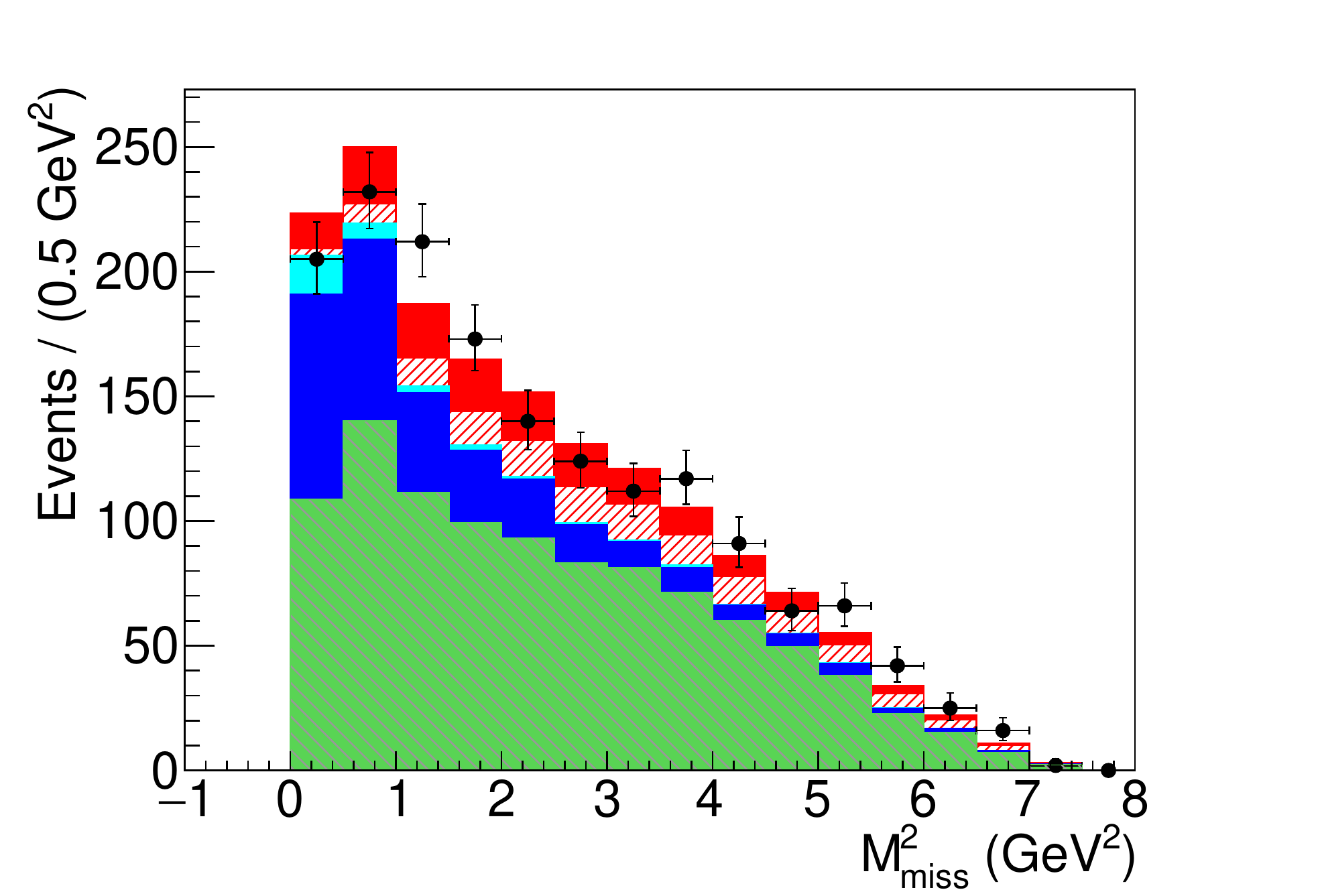}
  \includegraphics[width=7.5cm]{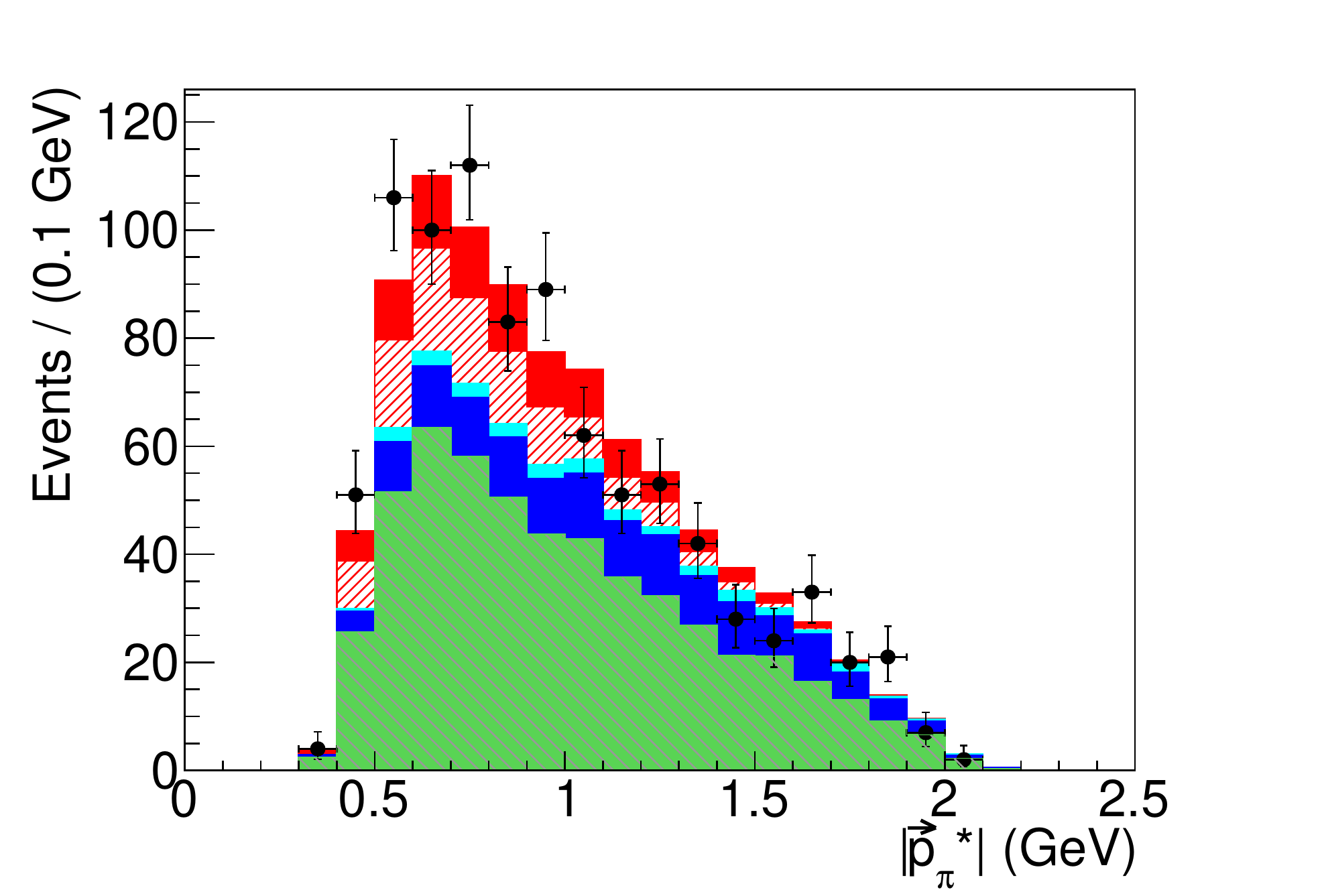}
  \includegraphics[width=7.5cm]{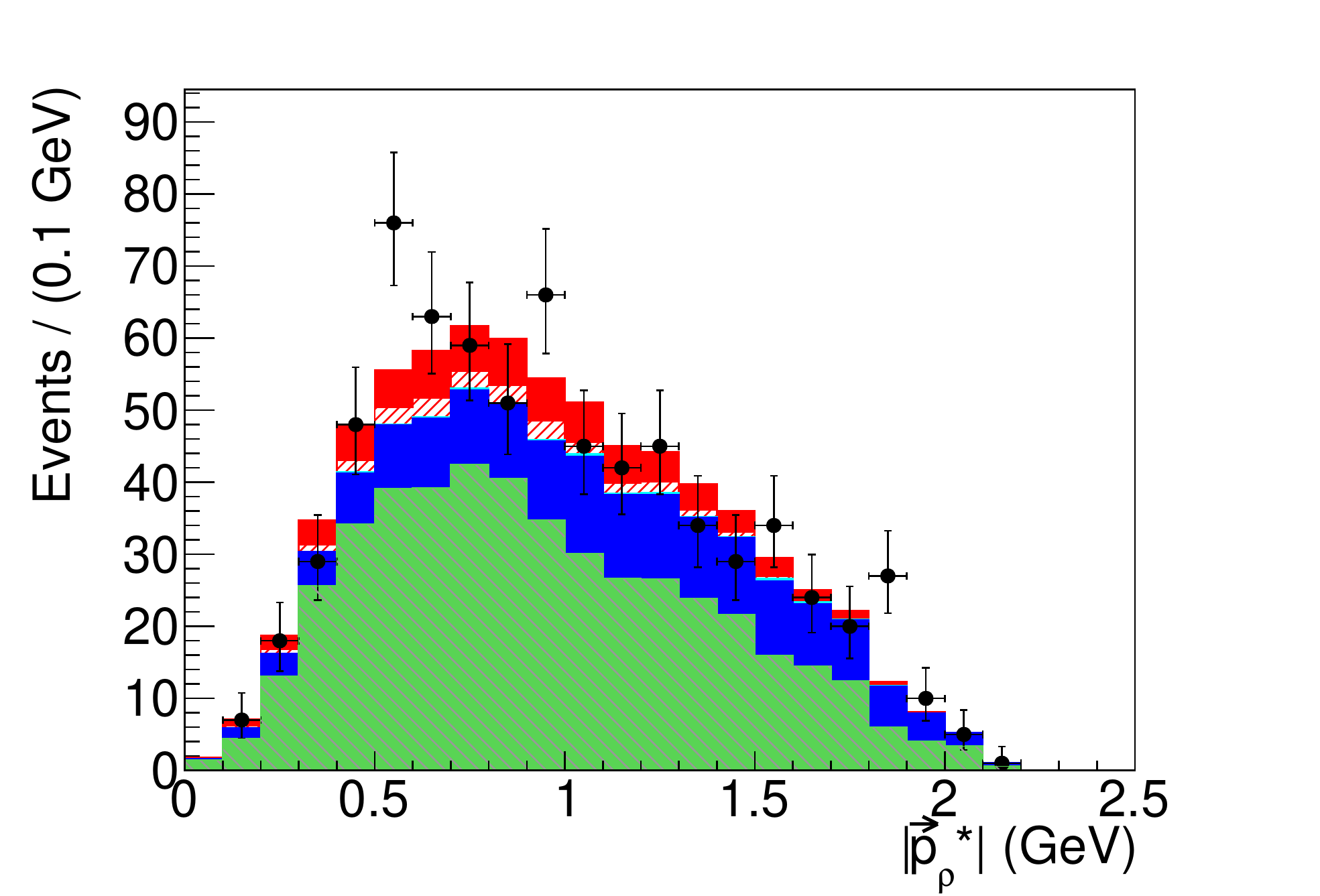}
  \caption{Projections of the fit results on the distributions of $q^2$ (top-left), $M_{\rm miss}^2$ (top-right), $|\vec{p}_\pi^{\kern2pt *}|$ (bottom-left) and $|\vec{p}_\rho^{\kern2pt *}|$ (bottom-right). These distributions are the sum of all the signal samples.}
  \label{fig:fitresult-projection}
\end{figure*}

The fit to the signal samples is performed as shown in Fig.~\ref{fig:fitresult}, with a $p$-value of 0.29. The signal yields for the charged and the neutral $B$ samples are $210 \pm 27$ and $88 \pm 11$, respectively, where the errors are statistical. The observables $R(D^*)$ and $P_\tau(D^*)$ are obtained using Eqs.~(\ref{eq:Rst}) and (\ref{eq:Pst}) for the correctly reconstructed signal events. The $\rho \leftrightarrow \pi$ cross feed yield is constrained by $R(D^*)$ and $P_\tau(D^*)$. The other cross feed yield is determined only by $R(D^*)$. The efficiency ratios for the correctly reconstructed signal events are $\epsilon_{\rm norm} / \epsilon_{\rm sig} = 0.97 \pm 0.02$ for the charged $B$ mode and $1.21 \pm 0.03$ for the neutral $B$ mode. The obtained results are
\begin{eqnarray}
  R(D^*)   &=& 0.270 \pm 0.035 ({\rm stat}),\\
  P_\tau(D^*) &=& -0.38 \pm 0.51 ({\rm stat}).
\end{eqnarray}

Figure~\ref{fig:fitresult-projection} shows the projections of the fit results in $q^2$, $M_{\rm miss}^2$, $|\vec{p}_\pi^{\kern2pt *}|$ and $|\vec{p}_\rho^{\kern2pt *}|$, where $\vec{p}_{\pi(\rho)}^{\kern2pt *}$ is the momentum of the $\tau$-daughter $\pi$ ($\rho$) in the CM frame. Each PDF component is scaled based on the yield obtained from the fit. All the panels show good agreement between the data and the expectation from the MC simulation.

\section{Systematic Uncertainties}\label{sec:syst}

\begin{table*}[t!]
  \centering
  \caption{The systematic uncertainties in $R(D^*)$ and $P_\tau(D^*)$, where the values for $R(D^*)$ are relative errors. The group ``common sources'' identifies the common systematic uncertainty sources in the signal and the normalization modes, which cancel to a good extent in the ratio of these samples. The reason for the incomplete cancellation is described in the text.}
  \vspace{3mm}
  \begin{tabular}{@{\hspace{0.5cm}}l@{\hspace{0.5cm}}|@{\hspace{0.5cm}}c@{\hspace{0.5cm}}@{\hspace{0.5cm}}c@{\hspace{0.5cm}}}
    \hline
    \hline
    Source & $R(D^*)$ & $P_{\tau}(D^*)$\\
    \hline
    Hadronic $B$ composition                 & $^{+7.7\%}_{-6.9\%}$ & $^{+0.134}_{-0.103}$\\
    MC statistics for PDF shape              & $^{+4.0\%}_{-2.8\%}$ & $^{+0.146}_{-0.108}$\\
    Fake $D^*$                               & 3.4\% & 0.018\\
    ${\bar B} \rightarrow D^{**} \ell^- {\bar \nu}_\ell$   & 2.4\% & 0.048\\
    ${\bar B} \rightarrow D^{**} \tau^- {\bar \nu}_\tau$   & 1.1\% & 0.001\\
    ${\bar B} \rightarrow D^* \ell^- {\bar \nu}_\ell$      & 2.3\% & 0.007\\
    $\tau$ daughter and $\ell^-$ efficiency  & 1.9\% & 0.019\\
    MC statistics for efficiency estimation  & 1.0\% & 0.019\\
    $\mathcal{B}(\tau^- \rightarrow \pi^- \nu_\tau, \rho^- \nu_\tau)$ & 0.3\% & 0.002\\
    $P_{\tau}(D^*)$ correction function      & 0.0\% & 0.010\\
    \hline
    \multicolumn{3}{c}{Common sources}\\
    \hline
    Tagging efficiency correction            & 1.6\% & 0.018\\
    $D^*$ reconstruction                     & 1.4\% & 0.006\\
    Branching fractions of the $D$ meson     & 0.8\% & 0.007\\
    Number of $B{\bar B}$ and $\mathcal{B}(\Upsilon(4S) \rightarrow B^+B^-$ or $B^0 \bar{B}^0)$ & 0.5\% & 0.006\\
    \hline
    Total systematic uncertainty & $^{+10.4\%}_{-9.4\%}$ & $^{+0.21}_{-0.16}$\\
    \hline
    \hline
  \end{tabular}
  \label{tab:syst-final}
\end{table*}

We estimate systematic uncertainties by varying each possible uncertainty source (such as the PDF shape and the signal reconstruction efficiency) with the assumption of a Gaussian error, unless stated otherwise. In several trials, we change each parameter at random, repeat the fit, and take the shifts of values of $R(D^*)$ and $P_\tau(D^*)$ from all such trials as the corresponding systematic uncertainty that is enumerated in Table~\ref{tab:syst-final}.

The most significant systematic uncertainty, arising from the hadronic $B$ decay composition, is estimated as follows. Uncertainties of each $B$ decay fraction in the hadronic $B$ decay background are taken from the measured branching fractions or estimated from the uncertainties in the calibration factors discussed in Sec.~\ref{sec:background}. The uncertainty of light meson resonances in the hadronic $B$ decays is taken into account by varying the fractions of these resonances within the maximum allowable range.

The limited MC sample size used in the construction of the PDFs is a major systematic uncertainty source. We estimate this by regenerating the PDFs for each component and each sample using a toy MC approach based on the original PDF shapes. The same number of events are generated to account for the statistical fluctuation.

The PDF shape of the fake $D^*$ component has been validated by comparing the data and the MC sample in the $\Delta M$ sideband region. However, a slight fluctuation from the decay ${\bar B} \rightarrow D \tau^- {\bar \nu}_\tau$ may have a significant impact on the signal yield since this component has almost the same shape as the signal mode, peaking at $E_{\rm ECL} = 0$~GeV. We incorporate an additional uncertainty by varying the contribution from the ${\bar B} \rightarrow D \tau^- {\bar \nu}_\tau$ component within the current uncertainties in the experimental averages~\cite{cite:PDG:2016}: $\pm 32\%$ for $B^- \rightarrow D^0 \tau^- {\bar \nu_{\tau}}$ and $\pm 21\%$ for ${\bar B^0} \rightarrow D^+ \tau^- {\bar \nu_{\tau}}$. We take the theoretical uncertainty on the $\tau$ polarization of the ${\bar B} \rightarrow D \tau^- {\bar \nu}_\tau$ mode into account, which is found to be 0.002 for $P_\tau(D^*)$ and negligibly small. In addition, we estimate a systematic uncertainty due to the small $M_{\rm miss}^2$ shape correction for the fake $D^*$ component, discussed in Sec.~\ref{sec:Mmiss2}. The systematic uncertainties related to the fake $D^*$ shape are 3.0\% for $R(D^*)$ and 0.008 for $P_\tau(D^*)$. The fake $D^*$ yield, fixed using the $\Delta M$ sideband, has an uncertainty that arises from the statistical uncertainties of the yield scale factors. The systematic uncertainties arising from the yield scale factors are 1.6\% for $R(D^*)$ and 0.016 for $P_\tau(D^*)$.

The uncertainty of the decays ${\bar B} \rightarrow D^{**} \ell^- {\bar \nu}_\ell$ are twofold: the indeterminate composition of each $D^{**}$ state and the uncertainty in the FF parameters used for the MC sample production. The composition uncertainty is estimated based on uncertainties of the branching fractions: $\pm 6\%$ for ${\bar B} \rightarrow D_1 (\rightarrow D^* \pi) \ell \bar{\nu}_\ell$, $\pm 12\%$ for ${\bar B} \rightarrow D_2^* (\rightarrow D^* \pi) \ell \bar{\nu}_\ell$, $\pm 24\%$ for ${\bar B} \rightarrow D'_1 (\rightarrow D^* \pi \pi) \ell \bar{\nu}_\ell$, and $\pm 17\%$ for ${\bar B} \rightarrow D_0^* (\rightarrow D^* \pi) \ell \bar{\nu}_\ell$ and $\pm 100\%$ for other modes and $\bar{B} \rightarrow D^{**} \tau^- \bar{\nu}_\tau$. We also estimate an uncertainty arising from the FF parameters in LLSW.

The uncertainties due to the FF parameters in the normalization mode ${\bar B} \rightarrow D^* \ell^- \bar{\nu}_\ell$ are estimated using the uncertainties in the world-average values~\cite{cite:HFLAV:2014}. In addition, the uncertainty arising from the $M_{\rm miss}^2$ shape correction for the normalization sample is estimated as an uncertainty related to ${\bar B} \rightarrow D^* \ell^- \bar{\nu}_\ell$.

The uncertainties on the reconstruction efficiencies of the $\tau$-daughter particles and the charged leptons arise from the particle identification efficiencies for $\pi^{\pm}$ and $\ell^{\pm}$ and the reconstruction efficiency for $\pi^0$. They are measured with control samples: the $D^{*+} \rightarrow D^0(\rightarrow K^- \pi^+)\pi^+$ sample for $\pi^{\pm}$, the $\tau^- \rightarrow \pi^- \pi^0 \nu_{\tau}$ sample for $\pi^0$, and the $\gamma \gamma \rightarrow \ell^+ \ell^-$ sample for charged leptons. The sample $J/\psi \rightarrow \ell^+ \ell^-$ from $B$ decays is also used in order to account for the difference in multiplicity between two-photon events and $B$ decay events.

Reconstruction efficiencies of the three ${\bar B} \rightarrow D^* \tau^- {\bar \nu}_\tau$ components: ``Signal'', ``$\rho \leftrightarrow \pi$ cross feed'' and ``Other $\tau$ cross feed'' are estimated using the signal MC sample. The efficiency uncertainties arising from the MC statistics are varied independently for each component.

Other minor uncertainties arise due to the branching fractions of the $\tau$ lepton decays and errors on the parameters of the $P_\tau(D^*)$ correction function.

In addition, common uncertainty sources between the signal sample and the normalization sample are estimated. Although they largely cancel in $R(D^*)$, there are some residual uncertainties from background components where yields are fixed based on MC expectation. Here, uncertainties on the number of $B{\bar B}$ and the branching fraction of $\Upsilon(4S) \rightarrow B^+ B^-, B^0 \bar{B}^0$ (1.8\%), tagging efficiencies (4.7\%), branching fractions of the $D$ decays (3.4\%), and $D^*$ reconstruction efficiency (4.8\%) are evaluated for their impact on the final measurements. For the $D^*$ reconstruction efficiency, the uncertainty originates from reconstruction efficiencies of $K_S^0$, $\pi^0$, $K^{\pm}$ and $\pi^{\pm}$, and is therefore correlated with the efficiency uncertainty of the $\tau$-daughter particles containing $\pi^{\pm}$ and $\pi^0$. This correlation is taken into account in the total systematic uncertainties shown in Table~\ref{tab:syst-final}.

\section{Result and Discussion}

Including the systematic uncertainty, we obtain the final results
\begin{eqnarray}
  R(D^*)   &=& 0.270 \pm 0.035 ({\rm stat}) ^{+0.028} _{-0.025} ({\rm syst}),\\
  P_\tau(D^*) &=& -0.38 \pm 0.51 ({\rm stat}) ^{+0.21} _{-0.16} ({\rm syst}),
\end{eqnarray}
with a signal significance of 7.1$\sigma$. The significance is taken from $\sqrt{2 \ln (L_{\rm max} / {L_0})}$, where $L_{\rm max}$ and $L_0$ are the likelihood with the nominal fit and the null hypothesis, respectively. The statistical correlation between $R(D^*)$ and $P_\tau(D^*)$ is 0.29, and the total correlation including systematics is 0.33.

\begin{figure}[t!]
  \centering
  \includegraphics[width=9cm]{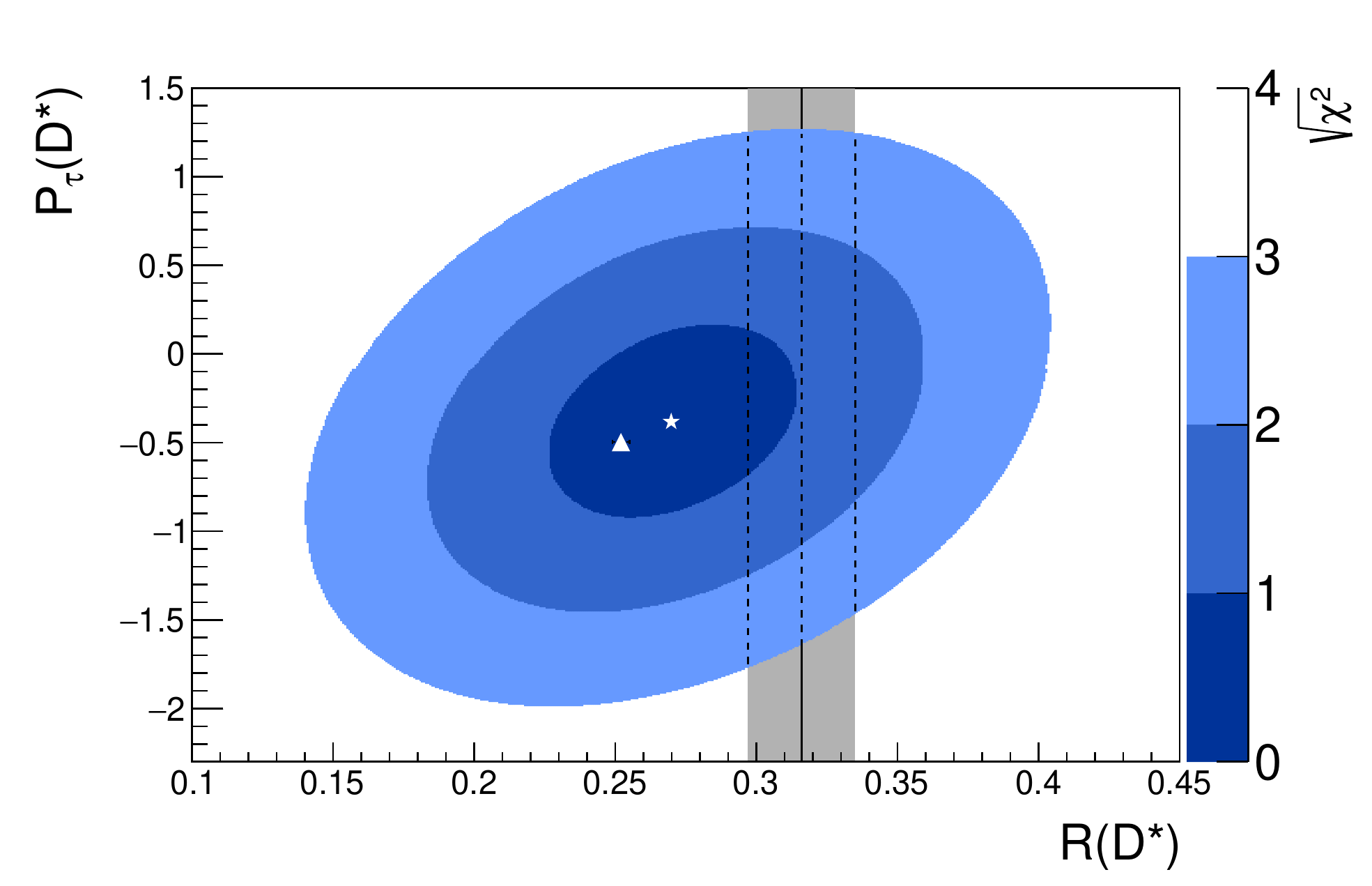}
  \caption{Comparison of our result (star for the best-fit value and $1\sigma$, $2\sigma$, $3\sigma$ contours) with the SM prediction (triangle). The white region corresponds to $>3\sigma$. The shaded vertical band shows the world average as of early 2016~\cite{cite:HFLAV:2014}.}
  \label{fig:ellipse}
\end{figure}

Figure~\ref{fig:ellipse} shows the exclusion region for the $R(D^*)$--$P_\tau(D^*)$ plane based on
\begin{eqnarray}
  \chi^2 &=& \left(
  \begin{array}{cc}
    \Delta^R & \Delta^P\\
  \end{array}
  \right)
  C^{-1}
  \left(
  \begin{array}{c}
    \Delta^R\\
    \Delta^P
  \end{array}
  \right),
\end{eqnarray}
where $\Delta^R = R(D^*) - 0.270$ and $\Delta^P = P_\tau(D^*) - (-0.38)$. The covariance matrix $C$ is represented by
\begin{eqnarray}
  C &=& \left(
  \begin{array}{cc}
    \left( \sigma_{\rm tot}^R \right)^2 & \rho_{\rm tot} \sigma_{\rm tot}^R \sigma_{\rm tot}^P\\
    \rho_{\rm tot} \sigma_{\rm tot}^R \sigma_{\rm tot}^P & \left( \sigma_{\rm tot}^P \right)^2\\
  \end{array}
  \right),
\end{eqnarray}
where $\rho_{\rm tot}$ and $\sigma_{\rm tot}^{R(P)}$ denote the total correlation factor and the total uncertainty on $R(D^*)$ [$P_\tau(D^*)$], respectively. Overall, our result is consistent with the SM prediction. Our measurement of $P_\tau(D^*)$ excludes the region larger than $+0.5$ at 90\% C.L.

\begin{figure}[t!]
  \centering
  \includegraphics[width=9cm]{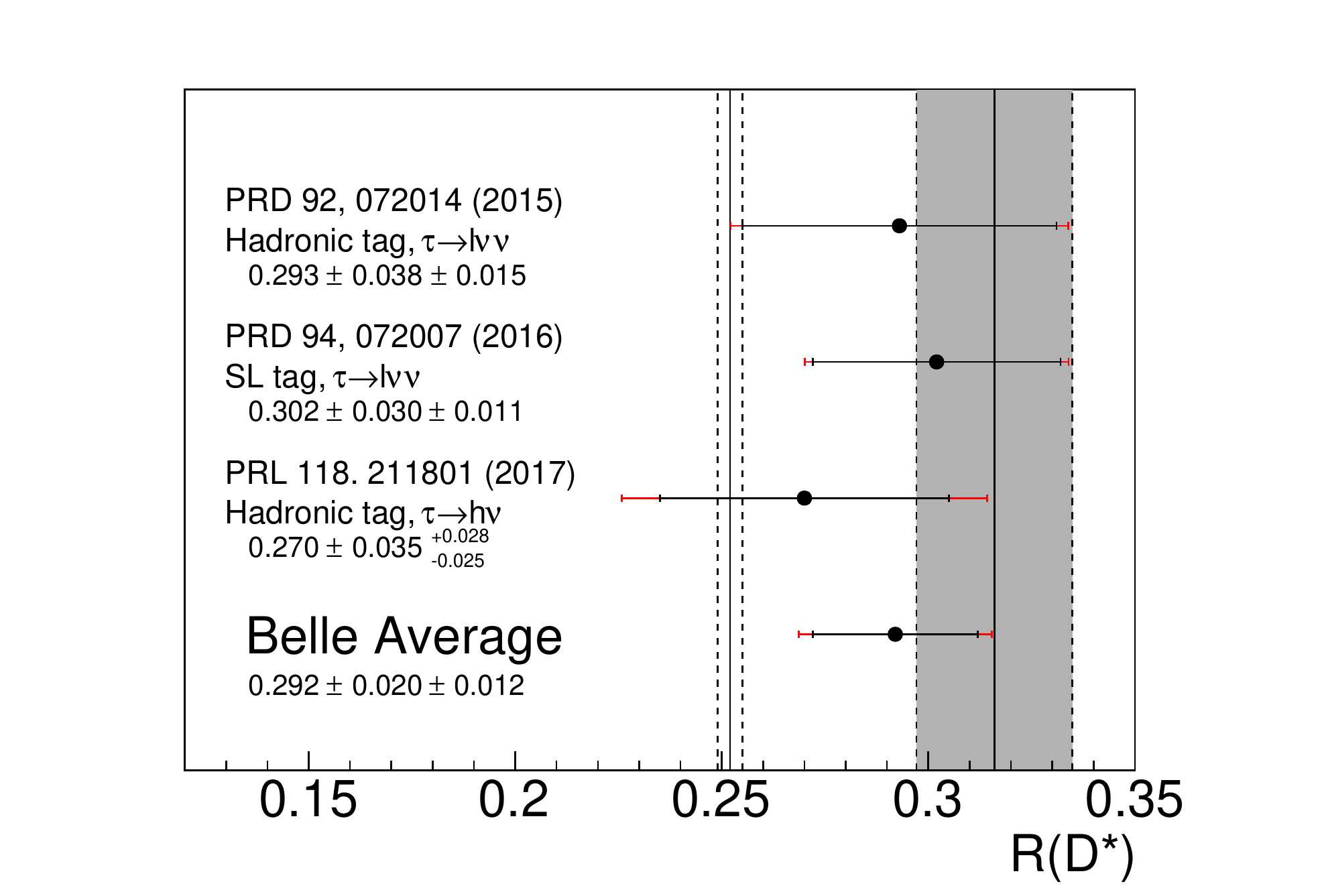}
  \caption{Summary of the $R(D^*)$ measurements based on the full data sample of Belle and their average. The inner (outer) error bars show the statistical (total) uncertainty. The shaded band is the world average as of early 2016~\cite{cite:HFLAV:2014} while the white band is the SM prediction~\cite{cite:RDst:2012}. On each measurement, the tagging method and the choice of the $\tau$ decay are indicated, where ``SL tag'' is the semileptonic tag and $h$ in the $\tau$ decay denotes a hadron $h = \pi$ or $\rho$.}
  \label{fig:average}
\end{figure}

As shown in Fig.~\ref{fig:average}, the obtained $R(D^*)$ also agrees with the previous Belle measurements: $R(D^*) = 0.293 \pm 0.038 \pm 0.015$~\cite{cite:Belle:2015} and $0.302 \pm 0.030 \pm 0.011$~\cite{cite:Belle:2016}, and with the world average as of early 2016~\cite{cite:HFLAV:2014}. Including our result and the latest LHCb result on $R(D^*)$~\cite{cite:LHCb:2017}, the world average is estimated to be $0.304 \pm 0.013 ({\rm stat}) \pm 0.007 ({\rm syst})$~\cite{cite:HFLAV:2016}.

The three results of $R(D^*)$ with the full data sample of Belle are statistically independent. The average $R(D^*)$ measured by Belle is estimated to be $0.292 \pm 0.020 ({\rm stat}) \pm 0.012 ({\rm syst})$. In this average, correlation in the uncertainties arising from background semileptonic $B$ decays is taken into account and other uncertainties are regarded as independent. The relative error in the average $R(D^*)$ is 7.5\%, which is the most precise result by a single experiment. Compared to the SM prediction~\cite{cite:RDst:2012}, the estimated value is 1.7$\sigma$ higher. Including $R(D)$ measured by Belle~\cite{cite:Belle:2015}, compatibility with the SM predictions is 2.5$\sigma$, corresponding to a $p$-value of 0.042.

\section{Conclusion}

We report the measurement of $R(D^*)$ with hadronic $\tau$ decay modes $\tau^- \rightarrow \pi^- \nu_{\tau}$ and $\tau^- \rightarrow \rho^- \nu_{\tau}$, and the first measurement of $P_\tau(D^*)$ in the decay ${\bar B} \rightarrow D^* \tau^- {\bar \nu_{\tau}}$, using $772 \times 10^6$ $B\bar{B}$ data accumulated with the Belle detector. Our results are
\begin{eqnarray}
  R(D^*)   &=& 0.270 \pm 0.035 ({\rm stat}) ^{+0.028} _{-0.025} ({\rm syst}),\\
  P_\tau(D^*) &=& -0.38 \pm 0.51 ({\rm stat}) ^{+0.21} _{-0.16} ({\rm syst}),
\end{eqnarray}
which are consistent with the SM predictions. The result excludes $P_\tau(D^*) > +0.5$ at 90\% C.L. This is the first measurement of the $\tau$ polarization in the semitaounic decays, providing a new dimension in the search for NP in semitauonic $B$ decays.

\begin{acknowledgements}

We acknowledge Y.~Sakaki, M.~Tanaka, and R.~Watanabe for their invaluable suggestions and help.

We thank the KEKB group for the excellent operation of the
accelerator; the KEK cryogenics group for the efficient
operation of the solenoid; and the KEK computer group,
the National Institute of Informatics, and the 
PNNL/EMSL computing group for valuable computing
and SINET5 network support.  We acknowledge support from
the Ministry of Education, Culture, Sports, Science, and
Technology (MEXT) of Japan, the Japan Society for the 
Promotion of Science (JSPS), and the Tau-Lepton Physics 
Research Center of Nagoya University; 
the Australian Research Council;
Austrian Science Fund under Grant No.~P 26794-N20;
the National Natural Science Foundation of China under Contracts
No.~11435013,
No.~11475187,
No.~11521505,
No.~11575017,
No.~11675166,
No.~11705209;
Key Research Program of Frontier Sciences, CAS, Grant No.~QYZDJ-SSW-SLH011;
the CAS Center for Excellence in Particle Physics (CCEPP);
Fudan University Grant No.~JIH5913023, No.~IDH5913011/003,
No.~JIH5913024, No.~IDH5913011/002;
the Ministry of Education, Youth and Sports of the Czech
Republic under Contract No.~LTT17020;
the Carl Zeiss Foundation, the Deutsche Forschungsgemeinschaft, the
Excellence Cluster Universe, and the VolkswagenStiftung;
the Department of Science and Technology of India; 
the Istituto Nazionale di Fisica Nucleare of Italy; 
National Research Foundation (NRF) of Korea Grants No.~2014R1A2A2A01005286, No.2015R1A2A2A01003280,
No.~2015H1A2A1033649, No.~2016R1D1A1B01010135, No.~2016K1A3A7A09005 603, No.~2016R1D1A1B02012900; Radiation Science Research Institute, Foreign Large-size Research Facility Application Supporting project and the Global Science Experimental Data Hub Center of the Korea Institute of Science and Technology Information;
the Polish Ministry of Science and Higher Education and 
the National Science Center;
the Ministry of Education and Science of the Russian Federation and
the Russian Foundation for Basic Research;
the Slovenian Research Agency;
Ikerbasque, Basque Foundation for Science and
MINECO (Juan de la Cierva), Spain;
the Swiss National Science Foundation; 
the Ministry of Education and the Ministry of Science and Technology of Taiwan;
and the U.S.\ Department of Energy and the National Science Foundation.
This work is supported by a Grant-in-Aid for Scientific Research (S) ``Probing New Physics with Tau-Lepton'' (No. 26220706) 
and was partly supported by a Grant-in-Aid for JSPS Fellows (No. 25.3096).

\end{acknowledgements}


\begin{thebibliography}{99}
\bibitem{cite:CC}
  Throughout this paper, the inclusion of the charge-conjugate mode is implied.
\end{thebibliography}
\end{document}